\documentclass[conference]{IEEEtran}
\IEEEoverridecommandlockouts

\usepackage{cite}
\usepackage{amsmath,amssymb,amsfonts}
\usepackage{algorithmic}
\usepackage{graphicx}
\usepackage{textcomp}
\usepackage{xcolor}
\usepackage[hyphens]{url}
\usepackage{fancyhdr}
\usepackage[bookmarks=true,breaklinks=true,letterpaper=true,colorlinks,citecolor=blue,linkcolor=blue,urlcolor=blue]{hyperref}
\usepackage{pifont}
\usepackage{newtxtext}
\usepackage{subcaption}
\usepackage[font=normalsize]{caption}
\usepackage{hyperref}
\usepackage{booktabs}
\usepackage{tabularx} 
\usepackage{multirow}

\newcommand{\ie}{\textit{i}.\textit{e}.}
\newcommand{\eg}{\textit{e}.\textit{g}.}

\newcommand{\xname}{Cambricon-LLM}
 
\def\BibTeX{{\rm B\kern-.05em{\sc i\kern-.025em b}\kern-.08em
    T\kern-.1667em\lower.7ex\hbox{E}\kern-.125emX}}
\begin{document}

\title{\xname: A Chiplet-Based Hybrid Architecture for On-Device Inference of 70B LLM\\
\thanks{$^{*}$Corresponding author}
\thanks{$^{\dagger}$Equal contribution}
}

\renewcommand{\IEEEauthorrefmark}[1]{%
    \textsuperscript{#1}%
}

\author{%
\IEEEauthorblockN{
    Zhongkai Yu\IEEEauthorrefmark{1,}\IEEEauthorrefmark{2}$^{\text{,}\dagger}$,
    Shengwen Liang\IEEEauthorrefmark{1}$^{\text{,}\dagger}$,
    Tianyun Ma\IEEEauthorrefmark{3},
    Yunke Cai\IEEEauthorrefmark{1,}\IEEEauthorrefmark{2},
    Ziyuan Nan\IEEEauthorrefmark{1,}\IEEEauthorrefmark{2},
    Di Huang\IEEEauthorrefmark{1},
    Xinkai Song\IEEEauthorrefmark{1},\\
    Yifan Hao\IEEEauthorrefmark{1},
    Jie Zhang\IEEEauthorrefmark{4},
    Tian Zhi\IEEEauthorrefmark{1},
    Yongwei Zhao\IEEEauthorrefmark{1},
    Zidong Du\IEEEauthorrefmark{1,}\IEEEauthorrefmark{5},
    Xing Hu\IEEEauthorrefmark{1,}\IEEEauthorrefmark{5}$^{\text{,}*}$,
    Qi Guo\IEEEauthorrefmark{1},
    Tianshi Chen\IEEEauthorrefmark{6}
}
\IEEEauthorblockA{\IEEEauthorrefmark{1}SKL of Processors, Institute of Computing Technology, CAS, Beijing, China}
\IEEEauthorblockA{\IEEEauthorrefmark{2}University of Chinese Academy of Sciences, Beijing, China}
\IEEEauthorblockA{\IEEEauthorrefmark{3}University of Science and Technology of China, Beijing, China}
\IEEEauthorblockA{\IEEEauthorrefmark{4}Peking University, Beijing, China}
\IEEEauthorblockA{\IEEEauthorrefmark{5}Shanghai Innovation Center for Processor Technologies}
\IEEEauthorblockA{\IEEEauthorrefmark{6}Cambricon Technologies Co., Ltd., China}
\IEEEauthorblockA{Emails: \{yuzhongkai21s, liangshengwen, caiyunke21e, nanziyuan21s, huangdi, songxinkai, haoyifan, zhitian, zhaoyongwei, \\
duzidong, huxing, guoqi\}@ict.ac.cn, mty21@mail.ustc.edu.cn, jiez@pku.edu.cn, tchen@cambricon.com}
}

\maketitle

\begin{abstract}

Deploying advanced large language models on edge devices, such as smartphones and robotics, is a growing trend that enhances user data privacy and network connectivity resilience while preserving intelligent capabilities. However, such a task exhibits single-batch computing with incredibly low arithmetic intensity,  which poses the significant challenges of huge memory footprint and bandwidth demands on limited edge resources. 

To address these issues, we introduce \xname, a chiplet-based hybrid architecture with NPU and a dedicated NAND flash chip to enable efficient on-device inference of 70B LLMs. Such a hybrid architecture utilizes both the high computing capability of NPU and the data capacity of the NAND flash chip, with the proposed hardware-tiling strategy that minimizes the data movement overhead between NPU and NAND flash chip. Specifically, the NAND flash chip, enhanced by our innovative in-flash computing and on-die ECC techniques, excels at performing precise lightweight on-die processing. Simultaneously, the NPU collaborates with the flash chip for matrix operations and handles special function computations beyond the flash's on-die processing capabilities. Overall, \xname~ enables the on-device inference of 70B LLMs at a speed of 3.44 token/s, and 7B LLMs at a speed of 36.34 token/s, which is over 22$\times$ to 45$\times$ faster than existing flash-offloading technologies, showing the potentiality of deploying powerful LLMs in edge devices.  

\end{abstract}


\section{Introduction}
Large Language Models (LLMs) have demonstrated remarkable performance across a variety of tasks, suggesting their potential to reshape labor markets and boost human productivity. As such potential unfolds, it is increasingly important to facilitate private LLM inferences on edge devices, such as robotics or even smartphones. By ensuring that everyone possesses a personalized LLM agent, we not only harness the transformative power of LLM agents but also cater to fundamental needs of privacy, customization, and accessibility.

However, the deployment of personal LLM agents on edge devices, where the predominant task type is single-batch computing, presents unique challenges compared to traditional LLM inference processes in the cloud. This mode of inference does not leverage inter-batch parallelism, exacerbating memory bandwidth bottlenecks and further restricting the deployment of LLMs on edge devices. Specifically, personal LLM deployment faced the following challenges:


\noindent $\bullet$ \textbf{Huge memory footprint on limited edge resources.} Research has consistently demonstrated that LLMs with a greater number of parameters exhibit enhanced emergent capabilities and superior performance\cite{opt}. 
However, the extensive parameter size presents considerable challenges in terms of memory, as exemplified by the Llama-70B model under INT8 quantization, which demands 70GB of memory. This requirement far exceeds the capacity of typical smartphone DRAMs. Furthermore, large parameter sizes inevitably lead to intensive data movement, which is the primary source of energy consumption during LLM single-batch inference on edge devices. Notably, the energy cost associated with moving a single bit of data is estimated to be 100-500$\times$ greater than that required for computation \cite{energy1, energy2, energy3, energy4}.

\begin{figure}[!h]
    \begin{subfigure}[t]{0.251\textwidth}
        \includegraphics[width=\textwidth]{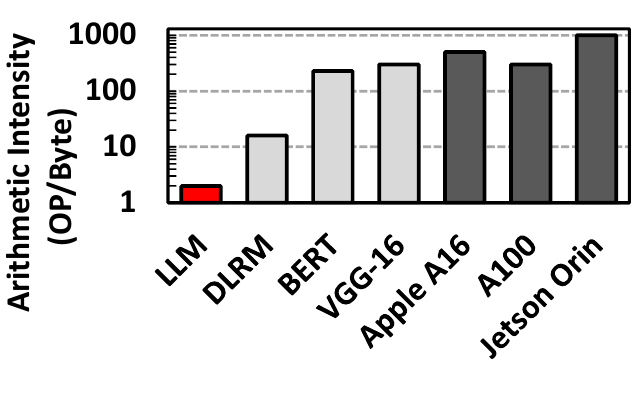}
        \vspace{-6mm}
        \caption{}
    \end{subfigure}
    \begin{subfigure}[t]{0.22\textwidth}
        \includegraphics[width=\textwidth]{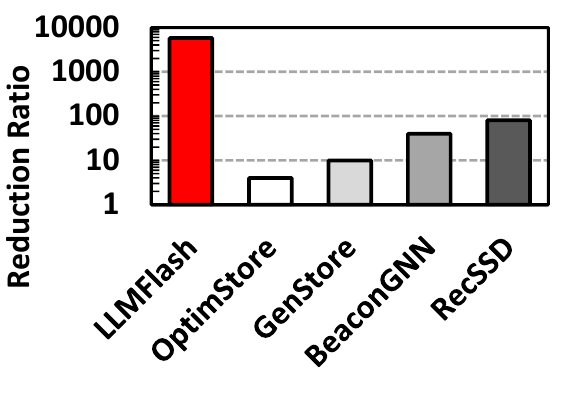}
        \vspace{-6mm}
        \caption{}
    \end{subfigure}
    \caption{\textcolor{black}{(a) Arithmetic Intensity comparison between LLM and other AI algorithms and (b) Reduction Ratio comparison between the scenario in LLMFlash and other ISC works.}}
    \label{fig:1_both}
      \vspace{-5mm}
  \end{figure}

\noindent $\bullet$ \textbf{Substantial bandwidth demand due to single-batch's incredibly low arithmetic intensity. }Arithmetic Intensity is defined as the ratio of the computational operations to the amount of data transferred between slow and fast memory. A Low arithmetic intensity suggests that the program is memory-bound. Unfortunately, single-batch LLM inference holds an unprecedentedly low arithmetic intensity of 2 under INT8 quantization. \textcolor{black}{As shown in \autoref*{fig:1_both}(a), the arithmetic intensity of LLM single-batch inference is 30$\times$ to 100$\times$ lower than that of other AI algorithms like DLRM~\cite{dlrm}, BERT~\cite{bert} and VGG\cite{vgg}, and over 100$\times$ lower than the capabilities of hardware, such as Apple A16, NVIDIA A100 and NVIDIA Jetson Orin.}
This gap leads to extremely low hardware utilization due to the pronounced memory access bottlenecks and an extremely high demand for bandwidth.

{To address the issue of the huge memory footprint,} several works such as FlexGen \cite{flexgen} and DeepSpeed \cite{DeepSpeed} have proposed offloading LLMs to flash-based SSDs. 
However, this approach has notable limitations. 
Firstly, the limited bandwidth of flash poses a critical bottleneck. For example, offloading the inference of INT8 quantized Llama2-70B model on UFS 4.0 flash (with 4GB/s bandwidth) yields a theoretical maximum speed of only 0.06 token/s, far from sufficient for real-time interactive applications that require a minimum of 3-10 token/s \cite{read_speed1, read_speed2,read_speed3,read_speed4_openai}.
Secondly, extra data transfer challenges the energy budget. 
Conventional architectures prevent NPUs from directly accessing data on flash, necessitating an initial data migration to DRAM, increasing the total data transfer by over 3$\times$. 
Given that data movement energy dominates the total energy consumption, these extra data transfers can result in significant energy overhead. {To meet the huge bandwidth demands,} in-storage computing (ISC) seems to be a potential solution by exploiting the inherent parallelism of flash. However, even the latest ISC works featuring on-die processing struggle to accommodate our scenario (\eg~OptimStore \cite{optimstore} and BeaconGNN \cite{beacongnn}).
Firstly, the high reduction ratio of LLM single-batch inference results in suboptimal flash channel utilization of under 10\%. 
The reduction ratio is defined as the ratio of input data size to output data size for a given operator. 
Taking the Llama2-7B model as an example, where over 95\% of computations involve general matrix-vector (GeMV) multiplications and the smallest matrix size is $4096 \times 4096$. 
After the computation, the result vector is reduced in size by a factor of 4096 compared to the original weight matrices. As shown in \autoref*{fig:1_both}(b), the reduction ratio of LLM single-batch inference is 100$\times$ larger than that observed in other scenarios discussed in previous ISC research, making them unsuitable.
Secondly, the absence of an on-die error correction mechanism in OptimStore and BeaconGNN leads to unreliable outcomes. 
If LLMs are exposed to flash memory errors without protection, their accuracy can decrease dramatically by over 70\%.

To this end, we propose \xname, a chiplet-based hybrid architecture with NPU and an on-die processing flash, to enable the efficient inference of 70B-LLMs on edge devices. 
As shown in \autoref*{fig:4_overall_arch}(a),  \xname~features a dedicated flash chip directly connected to the NPU via a chiplet die-to-die (D2D) link. This design effectively utilizes the data capacity of the flash chip and the computing capability of NPU, with the optimized tiling strategy that minimizes the data transfer between them. 
The integration of chiplet technology not only frees up flash from the bandwidth limitations of the UFS 4.0 protocol but also facilitates low-energy data transfers between NPU and flash, with an integrated flash controller for direct access from NPU to flash data. 
Furthermore, the flash memory within \xname~is equipped with on-die computation capabilities, allowing it to fully explore the inner bandwidth and reduce the data traffic on the flash channels.
In observing the reliability issue of in-storage computing architecture due to the frequent flash errors,  \xname~proposes an ultra-lightweight on-die Error Correction Unit to protect the outliers of model weights, thereby ensuring the accuracy of LLM inference.  


To the best of our knowledge, \xname~is {the first} hybrid architecture to extend NPU with a dedicated flash to achieve efficient single-batch inference of LLMs on edge devices.
We summarize the contributions of our work as follows:
\begin{itemize}
    \item We propose \xname, a novel chiplet-based hybrid architecture that combines an NPU and a specially designed flash.
    The flash is equipped with on-die processing capabilities and tailored for edge-based LLM inference.
    \item We introduce a hardware-aware tiling strategy that optimally distributes the LLM inference workload between NPU and flash, therefore fully using the available resources and avoiding flash channel underutilization.
    \item We propose an efficient on-die error correction algorithm and a lightweight hardware implementation, which effectively neutralizes the high error rates of flash memory, preserving the accuracy of LLM inference.
    \item We employ the SSDsim simulator \cite{ssdsim} to evaluate \xname~under various channel and chip configurations.
    \xname~achieves an inference speed of 3.44 tokens/s for 70B LLM, which is over $22\times$ faster than the baseline.
\end{itemize}
\section{Background}
\subsection{LLM Inference}
State-of-the-art LLMs \cite{gpt3, palm, palme, falcon, llama2,opt} are composed of multi-layered architectures where each layer corresponds to the decoder part of a Transformer model \cite{vaswani2017attention}. 
These LLMs feature an autoregressive computation pattern, where each output token is used as the input for generating the next token, leading to a sequential token generation process.
To minimize redundant computations of this process, the key and value vectors of previously generated tokens are stored for use in subsequent token computations, referred to as the KV cache. 
The size of the KV cache is proportionally related to both the batch size and the sequence length. 
In edge-based LLM inference, which typically uses a batch size of one, the KV cache remains small. For example, a 70B parameter LLM with a sequence length of 1000 would require a KV cache of around 700MB, a manageable size for the DRAM of edge devices, thus making on-device inference possible.

The inference process of Large Language Models (LLM) is divided into two distinct phases:

\noindent\textbf{The prefill phase} generates the first new token, \( x_{m+1} \), based on the \( m \) user-input tokens. 
Concurrently, the key and value vectors for the initial \( m \) tokens are computed and stored in DRAM as KV cache. 
Since all initial token values are known, these computations can occur in parallel, minimizing memory bottlenecks and allowing for efficient processing by existing NPU architectures.

\noindent\textbf{The decode phase} involves sequentially generating new tokens. For instance, to generate the \( (t+1)^{\text{th}} \) token, the model takes the \( t^{\text{th}} \) token as input and processes it through the layers to output the \( (t+1)^{\text{th}} \) token. 
Notably, the keys and values for all previous tokens have already been computed and stored, thus avoiding redundant calculations. \textcolor{black}{Although the generation of each new token has extremely low computational demand, it necessitates a whole transfer of weights to the chip, }presenting a significant memory bottleneck and posing the greatest challenge for LLM inference on edge devices. 
For example, for Llama-70B with INT8 quantization, the generation of a token requires only around 0.14 Tera operations but more than 70GB of memory access. 

\subsection{In-storage Computing Based on Flash}
\begin{figure}[!t]
    \vspace{-7pt}
    \centering
    \includegraphics[width=0.48\textwidth]{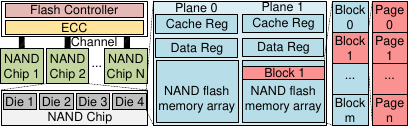}
    \caption{Architecture of Flash Memory.}
    \label{fig:2_flash}
\vspace{-7pt}
\end{figure}

\vspace{-0.8mm}
Flash memory is a high-density, non-volatile storage medium that provides smartphones with over 1 TB of storage capacity.
\autoref*{fig:2_flash} illustrates the primary architecture of flash memory. Within the flash, there are multiple independent channels, each externally connected to a corresponding Error-Correcting Code (ECC) module and flash controller. Each independent channel is equipped with several chips, sharing the bandwidth of the channel. In other words, only one chip can communicate with the channel at any given moment. The traditional hierarchy of flash memory includes channels, chips, dies, planes, blocks, and pages. The size of each page is 4, 8, or 16 KB, with each plane containing two registers of equivalent size to a page for data buffering. Read operations in flash memory are performed at the page level, whereas write operations are conducted at the block level. Write operations require a duration that is one to two orders of magnitude longer than read operations.

To break the bandwidth bottleneck of flash, numerous studies have moved computational modules to storage devices, known as in-storage computing (ISC). Currently, there are two main types of ISC technologies based on flash memory: on-controller computing \cite{ecssd, recssd2021,genstore2022,bio1_spectrum} and on-die computing \cite{optimstore, beacongnn, data3_deepstore}.
\noindent\textbf{On-controller computing} integrates specialized architectures and hardware resources at the flash controller side, allowing it to process computational tasks directly, and fully utilizing the bandwidth of all parallel flash channels. 
This design, commonly used in SSDs, helps overcome the limitations of external PCIe interface bandwidth.
For instance, while PCIe 4.0 SSDs have an interface bandwidth of only 8 GB/s, the total internal channel bandwidth can exceed 20 GB/s, making on-controller computing an effective solution.
However, this design does not address the issue of insufficient bandwidth within individual channels.

\noindent\textbf{On-die computing} refers to the addition of computational units within the die to reduce the volume of data transfer across channels. 
This design maximally exposes the parallel bandwidth of different chips/dies/planes to the computational units, thereby alleviating the pressure of data transmission on the channels. 
However, the limited space within the die restricts the complexity of computational logic that can be implemented. 
Additionally, moving computational units inside the die can render external ECC correction modules ineffective, potentially leading to computational errors. 
\section{Motivation}

In this section, we present the motivation of \xname~architecture. Section \ref{subsec:mot_lowari} emphasizes the unique characteristics of LLM inference tasks on edge devices. Section \ref{subsec:mot_why} justifies why flash memory is an ideal storage medium for this task.
In Section \ref{subsec:mot_ecc}, we conduct experiments to highlight the necessity of error-correction functionality for implementing high-accuracy LLM inference using flash memory.

\subsection{Low Arithmetic Intensity of Single-batch LLM Inference}\label{subsec:mot_lowari}
Arithmetic intensity quantifies the ratio of an algorithm's computational workload to its memory access volume.
A lower arithmetic intensity suggests a more pronounced memory-bound condition.
 \textcolor{black}{Remarkably, the decode phase of single-batch LLM inference under INT8 quantization shows an unprecedentedly low arithmetic intensity of 2.} 
This can be attributed to two primary reasons:
Firstly, the decoder-only architecture prevents prominent LLMs from exploring the parallelism across consecutive tokens, since tokens are generated sequentially rather than simultaneously.
Secondly, operating with a single batch size forces LLM to process only one request at a time, thereby eliminating the potential for batch-level parallelism.
These constraints significantly restrict the reuse of weight data across multiple tokens, thereby contributing to the extremely low arithmetic intensity observed in single-batch LLM inference.

\begin{figure}[!t]
    \centering
    \begin{subfigure}[b]{0.48\linewidth}
        \centering
        \includegraphics[width=1.0\textwidth]{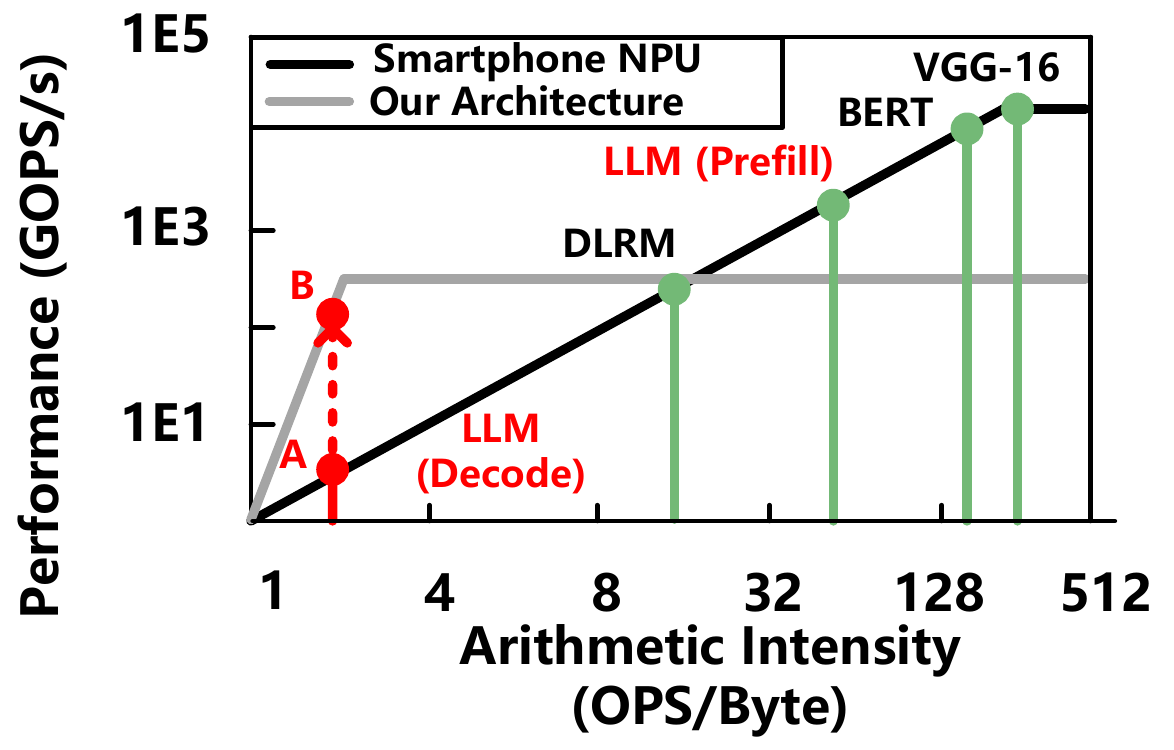}
        \caption{}
        \label{fig:3_roofline}
    \end{subfigure}
    \begin{subfigure}[b]{0.48\linewidth}
        \centering
        \includegraphics[width=1.0\textwidth]{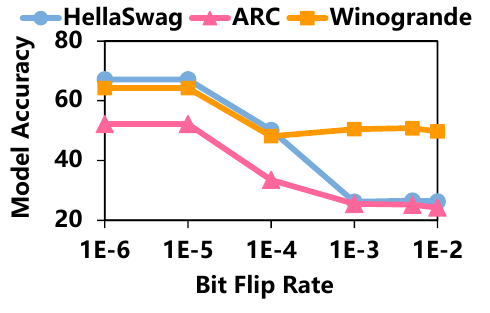}
        \caption{}
        \label{fig:3_flip}
    \end{subfigure}

    \caption{(a) Roofline model analysis of our architecture and smartphone NPU and (b) the sensitivity test of OPT-6.7B under bit-flip error in flash.}
    \label{fig:3_ecc_gate}
  \end{figure}

\autoref*{fig:3_roofline} illustrates the arithmetic intensity of different models \cite{arithmetic_cmu, arithmetic_bert}. 
Under INT8 quantization, the decode phase of LLM edge inference tasks has an Arithmetic Intensity of merely 2, which is orders of magnitude lower than that observed in the prefill phase and other model inference tasks. 
This leads to severe memory bandwidth bottlenecks when traditional mobile NPUs are employed. 
To enhance the efficiency of LLM inference at the edge, \xname~leverages on-die processing to minimize data transfer volume, and significantly improve the execution efficiency of the LLM inference during the decode phase, moving from point A to point B.

\subsection{Why Flash?}\label{subsec:mot_why}
Given the large parameter size, LLM inference can greatly benefit from the high storage density offered by flash memory. As indicated in \autoref*{table:3_density}, flash memory boasts a storage density of 10-30 Gb/mm$^2$, which is two orders of magnitude greater than that of DRAM \cite{flash_samsung_isscc2024,flash_skhynix_isscc2023,ddr5_isscc2023,lpddr5_isscc2024}. 
This makes flash the best choice for storing large models in space-constrained edge environments. 
Specifically, a typical 200GB NAND flash chip occupies about 64 mm$^2$, which is comparable to the area of a smartphone SoC, generally around 100 mm$^2$. 
This also proves the feasibility of the hybrid chiplet design proposed by \xname~from an area perspective.

Edge inference tasks are minimally impacted by the two primary disadvantages of flash memory in terms of read and write operations. 
\textcolor{black}{
\noindent{On one hand,} flash writes, which involvs programming and erasing, are executed in page-sized and block-sized units respectively. These operations are considerably slower than reads, often by several orders of magnitude.
However, for edge-based LLM inference tasks, which solely involve reading weight data from flash without any writing, this slow write speed becomes irrelevant.
\noindent{On the other hand,} data stored in flash is extracted from NAND flash memory array to data register in page-sized unit, ranging from 4 to 16KB.
}
This attribute can result in slow fragmented random data access when only a small portion of the page data is needed.
Nevertheless, this limitation is inconsequential for LLM edge inference, where data reads are sequential, extensive, and predictable.
For example, under INT8 quantization, even the smallest weight matrix of the llama2-7B model is 16MB, leading to the minimal overhead of splitting the model weights into page-sized segments.

\subsection{The Necessity of Error Correction}\label{subsec:mot_ecc}
Flash memory is known for its high error probability and the dominant are retention errors \cite{ssderror1_journal,ssderror2_journal}.
This error occurs due to electrons leaking from the floating gate of the floating-gate field-effect transistors, leading to alterations in the stored information. \textcolor{black}{The bit error rate of a new 3D TLC NAND chip can reach $1\times10^{-4}$~\cite{flash_error} after hours of retention time. As the flash ages with an increasing number of P/E cycles, the bit error rate can rise to over $1\times10^{-2}$~\cite{ssderror1_journal}.}
To mitigate these errors, SSDs are designed with complex error-correction logic in the controller, such as Low-Density Parity-Check (LDPC) error-correcting codes. 
However, due to the substantial area required by the error correction components, they cannot be integrated within the die, resulting in unavoidable error issues for on-die processing flash. 
\autoref*{fig:3_flip} demonstrates the effect of different error rates on the LLM inference outcome. 
These errors cannot be neglected, as they diminish the accuracy by over 70\%, making the result unreliable. 
Therefore, to ensure the accuracy of LLM inference, a lightweight error correction module is on demand for on-die processing flash. This module should be compact enough to fit within the die while still ensuring the precision of LLM inference.

\begin{table}[!t]
    \centering
    \caption{Storage density of DRAM and NAND Flash}
    \vspace{-1mm}
    \begin{tabular}{|c|c|c|c|}
    \hline
        Manufacturer & Type & Layers & Storage Density (Gb/mm$^2$) \\ \hline
        SK hynix & Flash & 300+ & 20.00 \\ \hline
        Samsung & Flash & 280 & 28.50 \\ \hline
        SK hynix & DDR & 1 & 0.30 \\ \hline
        SK hynix & LPDDR & 1 & 0.31 \\ \hline
    \end{tabular}
    \label{table:3_density}
\vspace{-12pt}
\end{table}
\section{\xname~Architecture}
This section presents the \xname~architecture.
Specifically, Section~\ref{subsec:overview} introduces the overview of \xname, Section~\ref{subsec:arch_die} presents the design of the Flash die, and Section~\ref{subsec:read_req_slice} describes the detailed design of the Slice Control.
\vspace{-12pt}
\subsection{Overview}
\label{subsec:overview}
\begin{figure*}[htbp]
    \centering
    \includegraphics[width=1.0\textwidth]{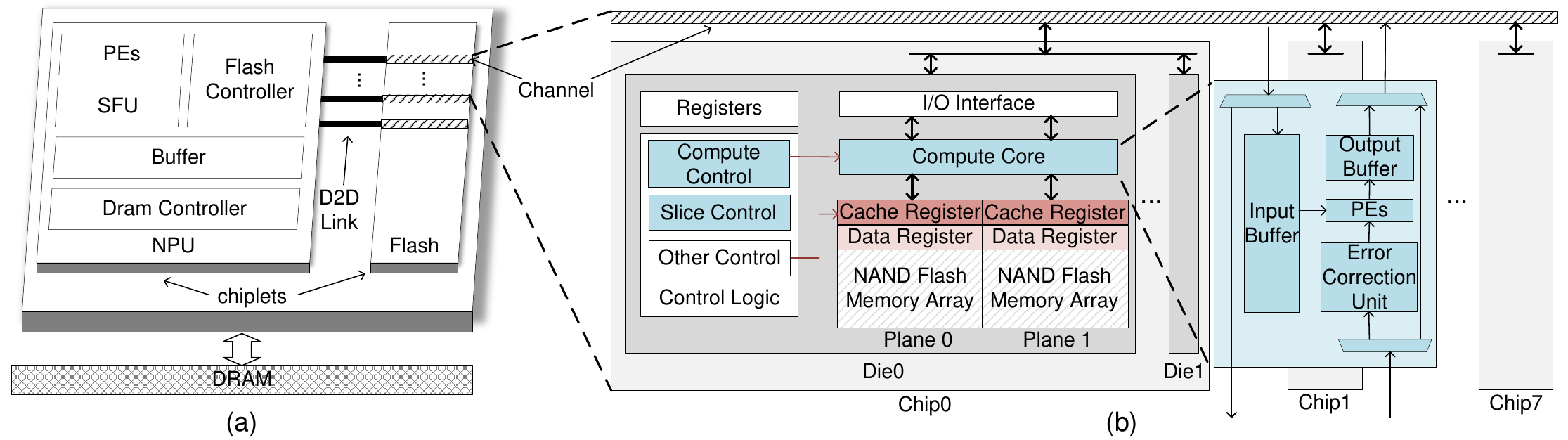}
    \caption{Overall architecture of \xname.}
    \label{fig:4_overall_arch}
\vspace{-12pt}
\end{figure*}

\autoref*{fig:4_overall_arch} (a) presents the overall \xname~architecture, which primarily consists of an NPU and a flash.
The flash equips on-die processing capabilities and connects to the NPU through high-speed Die-to-Die (D2D) chiplet links.
Additionally, the NPU integrates a flash controller to establish direct communication with the flash, marking a deviation from the designs of traditional NPUs.
Both the flash and the NPU contain Processing Elements (PEs) and work collaboratively to compute the GeMV multiplication between the weight matrices and the input vectors. 
\textcolor{black}{The NPU is equipped with a Special Function Unit (SFU) dedicated to managing specialized functions essential for LLM inference, such as Softmax, sin/cos, and ReLU. There is no SFU embedded in the flash as the on-die processing of these operations provides no tangible benefit and the complexity of implementing such logic on a flash die is prohibitively high.}



\begin{figure}[htbp]
    \centering
    \includegraphics[width=0.49\textwidth]{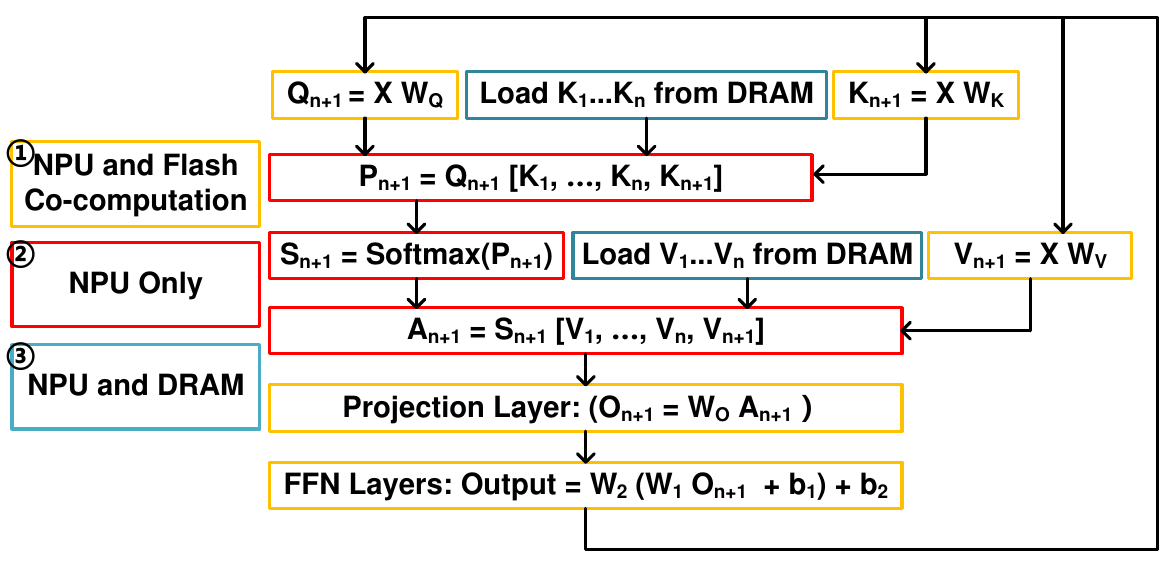}
    \caption{Compute flow and hardware mapping of an LLM inference layer.}
    \label{fig:4_compute_flow}
\vspace{-12pt}
\end{figure}

During single-batch inference of LLMs, the substantial and invariable weights remain in flash memory while the relatively small but frequently updated KV cache (\eg~less than 700MB for 70B LLM) is allocated to DRAM.
To efficiently process each layer of the LLM, we categorize all LLM operations into three groups and map them onto the hardware components of \xname~based on their unique characteristics. 
As shown in \autoref*{fig:4_compute_flow}, \ding{172} All GeMV operations that take the model weight as input are collaboratively executed by NPU and flash (yellow box). 
We utilize flash for these operations because its on-die processing units are positioned near the weight data and enable exploiting the internal parallelism of the flash.
We also incorporate the NPU to prevent underutilization of flash channels, which can occur from the high reduction ratio of these operations when using flash alone.
\ding{173} Matrix operations associated with the KV cache are exclusively handled by the NPU (red box). 
This is because the KV cache is stored in DRAM, which is closer to the PEs of the NPU.
\ding{174} Operations involving KV cache loading are managed jointly by the NPU and DRAM (blue box), simply because their functionality necessitates the involvement of both DRAM itself and the DRAM controller on the NPU.



To efficiently handle the GeMV multiplication with the flash die, we introduce a dedicated flash die design and a novel read-compute request, which will be discussed in Section \ref{subsec:arch_die}.



\subsection{Design of Flash Die}
\label{subsec:arch_die}
To enable the computation capability of the GeMV operations, each flash die in \xname~incorporates a novel shared Compute Core and two additional control logics (\ie, the Compute Control and the Slice Control), as depicted in \autoref*{fig:4_overall_arch} (b).
Note that though the shared Compute Core can only be exclusively occupied by a single plane during processing and increases the effort of the computational scheduling, it effectively mitigates the issues of significant area consumption and high error rates.
Specifically, the Compute Core is logically complex, encompassing multiple arithmetic units, buffers, and error correction modules. 
Thus, providing an independent Compute Core for each plane would lead to an unacceptable area consumption. 
Moreover, the Compute Core generates a considerable amount of heat during calculations, potentially raising the temperature of the flash memory array, which can in turn increase the error rate.
The increasing error rate could impact the inference accuracy when it surpasses the correction capabilities of the on-die Error Correction Unit.

Other than the Compute Core, the additional control logics in \xname~help enhance on-die processing capability as well. 
On the one hand, the Compute Control allows flexible execution of GeMV computations with diverse shapes.
On the other hand, the Slice Control plays a crucial role in effectively segmenting the entire read request into smaller slices, which prevents it from continuously occupying the channel and blocking subsequent requests. 
By efficiently managing the data flow, the Slice Control ensures smooth and uninterrupted processing. 
More details about the Slice Control will be discussed in Section \ref{subsec:read_req_slice}.

To better support the on-die computation, we propose a novel read-compute request instruction, supported by the Compute Core consisting of PEs, buffers, and an Error Correction Unit.
The read-compute request executes the GeMV multiplication inside each Compute Core as follows:
\ding{182} The input vector is sent to each Compute Core through the flash channel and stored in the input buffer of the Compute Core.
\ding{183} The weight matrix, originally stored in the NAND flash memory array, is fetched into the data register. 
Since the access of NAND flash memory array data is page-based, the size of the weight matrix for each computation must be the same as the page size.
\ding{184} The weight matrix is then transferred from the data register to the cache register to be used for further computation, leaving the data register empty to serve the next request, and thus form a pipeline format. 
\ding{185} The PEs in the Compute Core conduct GeMV multiplication using the matrix from the cache register and the vector from the input buffer. 
The resulting vector is stored in the output buffer.
Notably, the computing power of the Compute Core must match the read speed of the flash memory array.
For example, for a certain flash that requires 20us (\textit{tR}) to read a 16KB (page size) weight matrix, the PE must complete 32K operations (INT8) within 20us to avoid throughput delays.
\textcolor{black}{This corresponds to a computing power of 1.6 GOPS, which needs approximately two Multiply-Accumulate units (MAC).}
\ding{186} When the computation is finished, the result vector is sent back to the NPU through the flash channel. 

\subsection{Design of Slice Control}
\label{subsec:read_req_slice}
To enhance channel bandwidth utilization, \xname~introduces a novel Slice Control within the die, aiming to maximize the utilization of the available bandwidth. The Slice Control achieves this goal in two steps.
\textbf{First,} the result vector of the read-compute request has a relatively small size compared to the original weight matrix, and their transfer time through the flash channel is negligible compared to the page read time \textit{tR}, causing a low bandwidth utilization ($\le$ 6\%).
To address the low bandwidth utilization with only read-compute requests, the idle plane serves normal read requests to deliver model weights to the NPU when the other plane processes read-compute requests, avoiding the waste of limited channel bandwidth.
\textbf{Second,} our experiments reveal that transferring read-request data continuously in a page manner through the flash channel is suboptimal, as the prolonged transfer time can not fit in the channel occupancy bubbles between read-compute requests, leading to the blocking of subsequent read-compute requests. 
To mitigate this problem, we introduce a novel read request slice mechanism that segments a page-sized read request into smaller slices. 
The read requests are transferred slice by slice during the channel occupancy bubbles until the entire page is transferred so that the channel resources can be better used and the blocking of subsequent requests no longer exists. 


\begin{figure}[!t]
    \centering
    \includegraphics[width=0.5\textwidth]{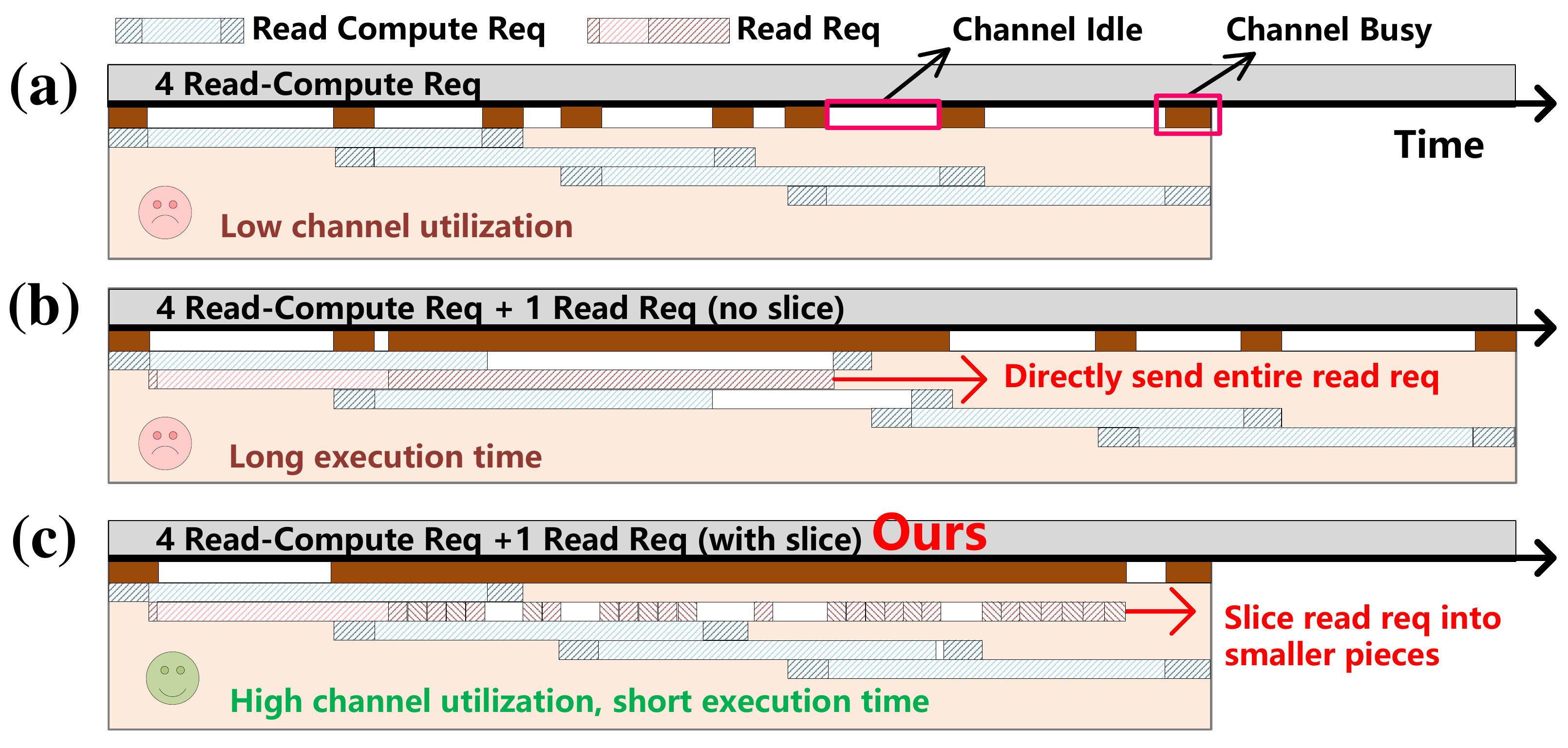}
    \caption{An example of the pipeline of the requests and the channel state with the three Slice Control strategies, (a) with only read-compute requests, (b) with both read-compute requests and read requests, and (c) with both read-compute requests and sliced read requests (ours).}
    \label{fig:4_slice_data_trans}
    \vspace{-12pt}
\end{figure}

Take a simplified configuration as an example, in which one channel connects one chip, with a single die that contains two planes and a shared Compute Core.
\autoref*{fig:4_slice_data_trans} shows the pipeline of the requests and the channel state with the three Slice Control strategies, \ie, with only read-compute requests (a), with both read-compute requests and read requests (b), and with both read-compute requests and sliced read requests (c). 
{For strategy (a)}, the continuous issuance of four read-compute requests leads to suboptimal bandwidth utilization, evidenced by the extensive white spaces in the black bars. This inefficiency stems primarily from the mismatch between the flash memory data read times and the result vector transfer times.
{For strategy (b)}, one unsliced read request interleave with four read-compute requests, resulting in severe blocking of the last two read requests and an extended total execution time. 
{For strategy (c)} with the proposed read request slice mechanism, the sliced read requests are interposed among four read-compute requests. This arrangement eliminates the blocking of subsequent read-compute requests and therefore significantly improves the channel utilization, making the execution time align with (a).

\section{Hardware-aware Tiling}
To optimize the use of available flash channel bandwidth resources, we propose a novel hardware-aware tiling strategy that efficiently allocates the LLM inference workload between the NPU and the flash.
This approach sophistically partitions the GeMV workload across both the NPU and the flash, and utilizes the idle periods of the flash channel when processing the read-compute requests to transfer a portion of the weight matrix to the NPU for processing. 
The hardware-aware tiling strategy works in two steps. 
\textbf{Firstly}, we identify the optimal tile shapes that align best with the hardware specifications of the flash (e.g. page size, channel number, and chip number).
\textbf{Secondly}, we determine the optimal workload distribution proportions to balance the computation between the NPU and the flash.


\begin{figure}[!t]
    \centering
    \includegraphics[width=0.45\textwidth]{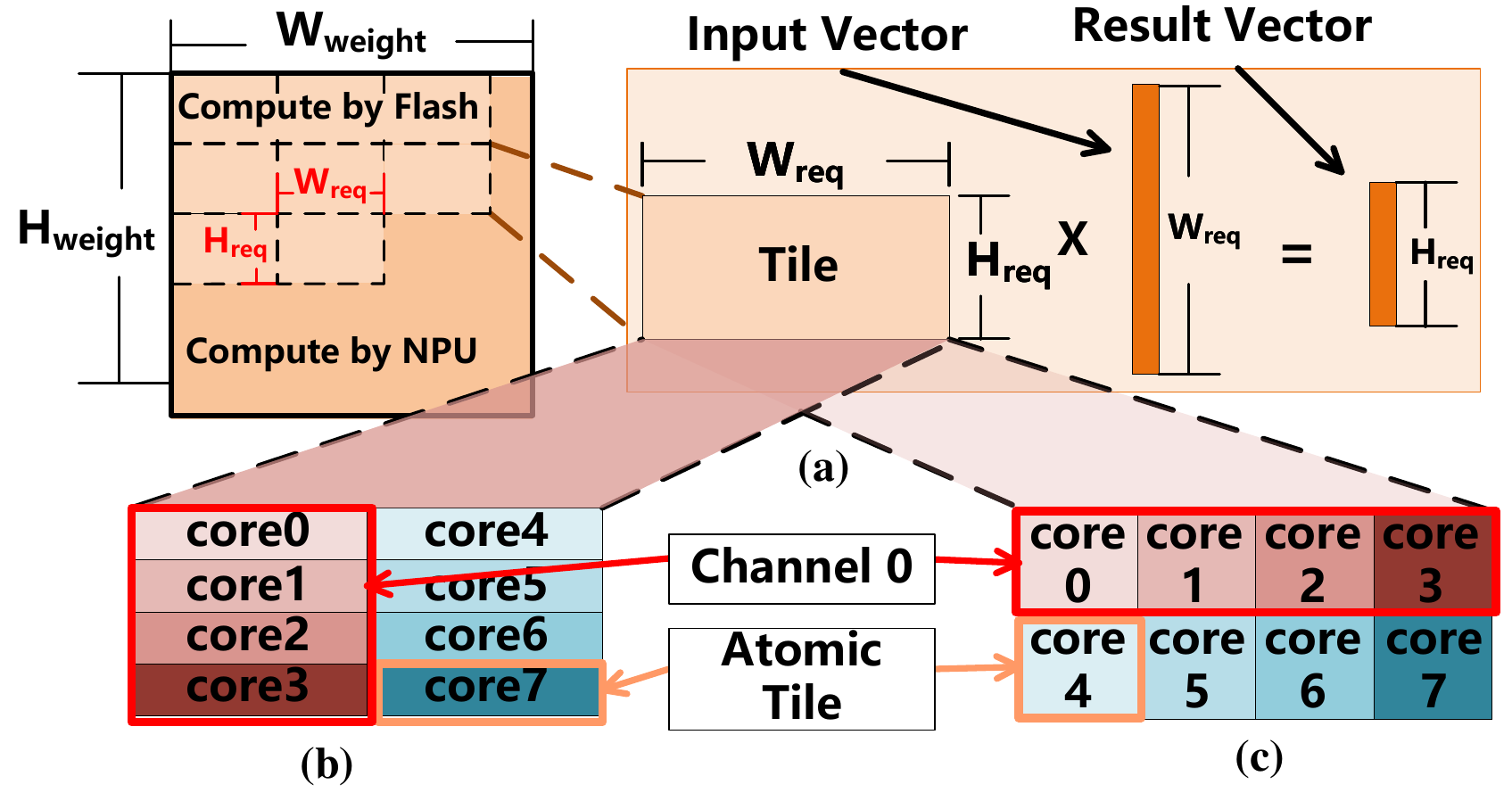}
    \caption{The hardware-aware tiling strategy in \xname.}
    \label{fig:5_split_matrix}
\vspace{-12pt}
  \end{figure}

\subsection{Identify the Optimal Shape of Tiles}
We first identify the optimal shape of the matrix tiles which match the hardware specifications of the flash.
As shown in \autoref*{fig:5_split_matrix} (a), during LLM inference, the whole weight matrix ($H_{\text{weight}} \times W_{\text{weight}}$) is segmented into smaller tiles ($H_{\text{req}} \times W_{\text{req}}$).
Each tile is processed collaboratively by all Compute Cores in the flash, corresponding to a single read-compute request.
The efficiency of LLM inference heavily relies on the shape of shapes, as it determines the lengths of input and result vectors, consequently impacting the overall volume of data transmitted through the flash channels.
Our goal is to identify the shape of the tile to minimize the overall data transfer.

Consider a simple flash setup with 2 channels, each connected to 4 Compute Cores (cores 0-3 on channel 0, and cores 4-7 on channel 1).
To fully utilize all computational resources in the flash, each tile will be evenly distributed to all 8 Compute Cores, with each Compute Core handling a segment named atomic tile, as described in \autoref*{fig:5_split_matrix} (b). 
Each colored block represents an atomic tile computed by a specific Compute Core, and blocks of the monochromatic color indicate allocation to the same channel.
Define $channel_{num}$ as the total number of channels in the flash, and $ccore_{num}$ as the total number of Compute Cores connected to each channel.
The tile is split vertically by channels into 2 segments, with each channel handling a segment of shape ($H_{req} \times \frac{W_{req}}{channel_{num}}$).
Given that $ccore_{num}$ Compute Cores are connected to each channel, each Compute Core handles an atomic tile of shape ($\frac{H_{req}}{ccore_{num}} \times \frac{W_{req}}{channel_{num}}$).

Each atomic tile is independently computed by one Compute Core and the total data transfer volume through the flash channel equals the size of the input vector plus the size of the result vector. 
Given $channel_{num} \times ccore_{num}$ Compute Cores on the flash, we can get a naive data transfer volume of $Trans = ccore_{num} \times W_{req} + channel_{num} \times H_{req}$. 
However, we observe that the four cores on the same channel (\eg core 0/1/2/3) share mutual input vectors. 
Therefore, \textbf{broadcast operations} can be employed to simultaneously send input vectors to all Compute Cores on the same channel. 
Those input vectors are stored in the input buffer of each Compute Core for further computation. This avoids repeatedly sending the same input vector multiple times through the flash channel and therefore reduces the total transferred data volume to:
$$Trans = W_{req} + channel_{num} \times H_{req}\text{.}$$

After deriving the expression for the total data volume transferred, our next objective is to determine the optimal tile dimensions ($H_{\text{req}}$ and $W_{\text{req}}$) that minimize $Trans$.
Given that the size of each atomic tile matches the page size, the following relationship is established under INT8 quantization:
$$H_{\text{req}} \times W_{\text{req}} = channel_{num} \times ccore_{num} \times pagesize\text{.}$$
Consequently, the formula for $Trans$ represents a standard AM-GM inequality, allowing us to readily compute the minimal data transfer size:
\begin{align*}
  \min\{Trans\} &= 2 \times \sqrt{W_{req} \times H_{req} \times channel_{num}} \\
  & = 2 \times channel_{num} \times \sqrt{ccore_{num} \times pagesize}\text{.}
\end{align*}
This minimum is attained when:
\begin{align*}
    H_{req}* &= \sqrt{ccore_{num} * pagesize} \text{,}\\
    W_{req}* &= channel_{num} * \sqrt{ccore_{num} * pagesize}\text{.}
\end{align*}
For now, we identify the optimal tile shape for the flash, which depends solely on the number of channels, the number of Compute Cores, and the page size.
During LLM inference, we tailor each weight matrix into this specific shape to maximize the inference speed.

It is important to note that this analysis is based on the splitting scheme shown in \autoref*{fig:5_split_matrix} (b)
There is, however, an alternative splitting scheme illustrated in \autoref*{fig:5_split_matrix} (c).
In this alternative approach, the input vector of atomic tiles shows no chance of data reuse, resulting in a total data transfer volume of: 
$Trans_2 = core_{num} \times W_{req} + channel_{num} \times H_{req}$
This expression also follows the AM-GM inequality, but the minimum, $2 \times \sqrt{W_{\text{req}} \times H_{\text{req}} \times channel_{\text{num}} \times ccore_{\text{num}}}$, is larger than $\min\{Trans\}$, rendering this scheme less optimal.

.

\vspace{-12pt}
\subsection{Determine Optimal Workload Distribution Proportions}
After the optimal tiling shape is determined, we then need to determine the proportion of workload allocated to NPU and flash, respectively.
The primary goal is to balance the execution times of the flash and the NPU.
While the execution time on the flash depends on the read-compute requests, the execution time on the NPU is dominated by data transfer time (read requests), given its considerable computing power and the modest computational demand of LLM single-batch inference. 
Therefore, optimizing the workload distribution requires a careful balance between the execution times of read-compute requests and read requests.
To achieve this, we first estimate the execution time for a single read-compute request and a single read request, then determine the proportion that balances the total execution time for all read-compute requests and read requests.


For read-compute requests, neglecting pipeline startup latency, the execution time $t_{rc}$ is the sum of the input data transmission times through the flash channel and the read operation latency from the flash memory array:
$$t_{rc} = t_{R} + \frac{W_{req}}{channel_{num} \times bw_{channel}}\text{.}$$
Where $t_{R}$ is the read operation time, $W_{req}$ is the tile width, $channel_{num}$ is the number of channels, and $bw_{channel}$ is the channel bandwidth.
To determine the execution time of read request, we first calculate the channel utilization rate for read-compute requests, which informs us how much bandwidth remains for the read requests. 
\textcolor{black}{
$$rate_{rc} = \frac{H_{req} + \frac{W_{req}}{channel_{num}}}{t_{R} \times bw_{channel}}\text{.}$$}
Then, the execution time for a read request ($t_{r}$) can be estimated as:
$$t_{r} = \frac{pagesize}{(1-rate_{rc}) \times bw_{channel}}\text{.}$$
The optimal workload distribution is achieved when the execution times for read and read-compute requests are equal. 
Denote the proportion for read compute workload as $\alpha$, then:
$$\alpha = \frac{t_{r}}{t_{r} + t_{rc}}\text{.}$$

Based on the distribution proportion $\alpha$, and the optimal tile shape $H_{req}$ and $W_{req}$, we devise a tiling strategy for all GeMV multiplication operations during LLM inference. 
As depicted in \autoref*{fig:5_split_matrix} (a), an $\alpha$ proportion of the weight matrix ($H_{weight} \times W_{weight}$) is assigned to flash in a tiled manner, each tile sized ($H_{req} \times W_{req}$).
In this figure, there are 8 tiles being allocated to flash in total and the remainder weight matrix is transferred through flash channels and processed on the NPU.

\section{Error Correction Mechanism}
To address the high error rates of the flash memory and ensure accurate LLM inference, we propose an efficient on-die error correction algorithm and a lightweight hardware implementation.
\textbf{Our key observation} is that when dealing with LLM models containing over 7 billion (7B) parameters, the model's accuracy is particularly affected by a small subset of outliers, which account for less than 0.1\% of the total parameter values \cite{quant1_olive, outlier_llmint8}.
These outliers have significantly larger absolute magnitudes compared to the regular parameter elements.
Based on this insight, we develop an outlier-oriented error correction code (ECC) and a hardware-friendly algorithm to enable on-die error correction.
The ECC is stored in the spare area of flash memory that is originally designated for ECC storage and occupies 1664 bytes for a 16KB page.

The error correction operates on two fronts.
Firstly, it safeguards outliers in the original weight matrix against flash errors.
Secondly, it prevents normal values from being transformed into outliers due to bit-flip errors in flash memory. 
For the outliers protection, we identify the top 1\% largest values within the entire page, and then store their positions and N copies of their values into the error correction code (ECC) for protection (N is an even number). 
During the inference of LLM, the ECC is decoded on die and a bit-wise majority vote is conducted between the N copies and the original data.

\begin{figure}[!t]
  \centering
  \begin{subfigure}[b]{0.40\linewidth}
      \centering
      \includegraphics[width=1.0\textwidth]{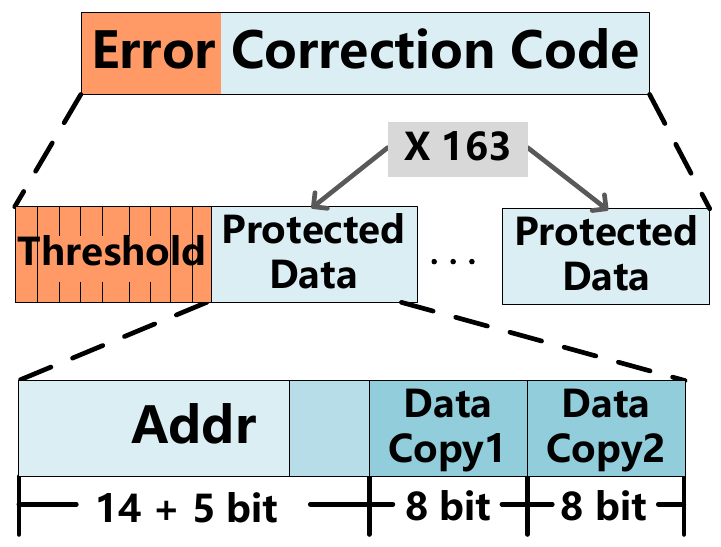}
      \caption{}
      \label{fig:6_err_code}
  \end{subfigure}
  \begin{subfigure}[b]{0.56\linewidth}
      \centering
      \includegraphics[width=1.0\textwidth]{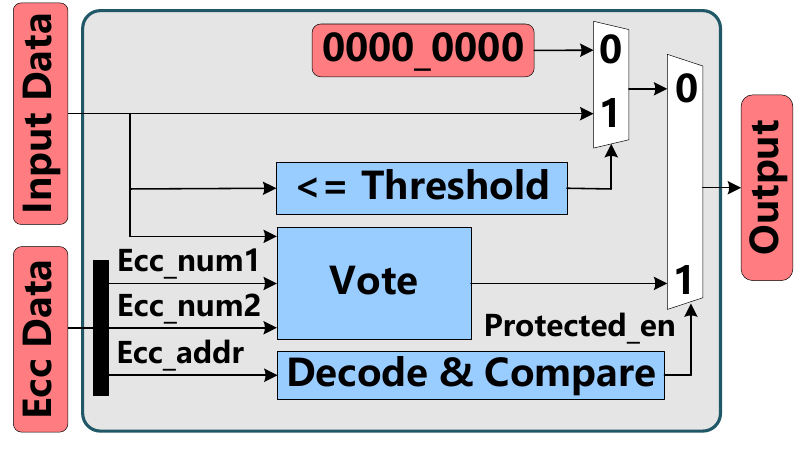}
      \caption{}
      \label{fig:6_gate_design}
  \end{subfigure}

  \caption{(a) The structure of our proposed ECC and (b) the design of on-die error correction unit in \xname.}
  \vspace{-5mm}
  \label{fig:6_ecc_gate}
\end{figure}

Take the example of an outlier valued 64(0b'0100\_0000) and $N=2$ (meaning the ECC contains 2 copies of the outlier value). 
If the three recorded instances are 01\textcolor{red}{1}0\_0000, 0100\_0000, and 0100\_0000 respectively, then the output value after decoding will be 0100\_0000, thus preserving the original outlier data under flash memory error.
TO quantify the protection capability of this mechanism, we set the single-bit flip rate of the flash array to be $x$. Only if more than $N/2+1$ of the $N+1$ instances of the outlier copies are flipped, will the output be flipped.
Therefore, the flip rate of protected outliers can be diminished to:
\begin{align*}
  f_{prot} &= \sum_{i=N/2+1}^{N+1} \binom{N+1}{i} \times x^{i}(1-x)^{N+1-i}  \\
    &\approx \binom{N+1}{N/2 + 1} \times x^{N/2 + 1} \\
    &= \frac{(N+1)!}{(N/2+1)! \times (N/2-1)!} \times x^{N/2 + 1}\text{.}
  \end{align*}
If $N=2$ and the original flip rate is $1\times 10^{-4}$, then the protected flip rate $f_{prot} = 3 x ^2 = 3\times 10^{-8}$.

To prevent normal values from being affected, while encoding the top 1\% largest values of the page, we also record the smallest among these large values as a threshold. 
During decoding, if any value exceeds the threshold yet is not marked in the ECC as an outlier, it is inffered to be generated by flash memory errors. 
To prevent such fake large values from compromising the accuracy, we trunk these values to zero.

Additionally, we propose the detailed ECC structure and a lightweight on-die error correction hardware implementation to  process the algorithm.
As shown in \autoref*{fig:6_err_code}, the ECC data structure stores the threshold at the beginning for multiple copies (e.g., 9 copies) to ensure its safety since it is crucial.
Following the threshold, the 1\% protected values in the entire page are stored.
For a 16KB page with the weight format as INT8, there are a total of 16384 data points in a page, thus a total of 163 large values are protected. 
For each protected large value, its address is stored along with two copies of the data. 
When each page contains 16,384 elements, a total of 14 bits are required to describe the address of each element. 
As the address itself can also be prone to errors, each address is accompanied by a 5-bit private error-correcting code for their safety, utilizing the format of Hamming code. 
This encoding format has a low hardware decoding overhead and can be efficiently completed on-die. 
If a 1-bit error occurs in the address, it will be corrected by the on-die decoder. 
If a 2-bit error occurs, the protected value will be discarded and treated as unprotected. 
The total size of the ECC for each page is $8\times 9 + (14+5 + 8\times 2) \times 163 = 722\text{B}$, which is less than the size of the spare space for each page (1664B).

\autoref*{fig:6_gate_design} depicts the hardware implementation of the on-die Error Correction Unit.
For each data retrieved from flash memory, it first evaluates whether the current input data falls within the protected range by comparing the address of current input data to the address of protected outliers stored in ECC. 
If the input data is identified as protected, error correction is enabled through a voting mechanism utilizing the two copies stored in ECC along with the input data itself. 
Otherwise, if the input data is identified as unprotected, its value should be no larger than the threshold.
Following this principle, the on-die Error Correction Unit compares the unprotected data with the threshold and directly output it if the unprotected data is smaller than the threshold.
Otherwise, this unprotected data is deemed an anomalous value likely resulting from flash memory error and is trunked to zero.
In summary, with the efficient on-die error correction algorithm and its lightweight hardware implementation, \xname~effectively neutralizes the high error rates of flash memory, preserving the accuracy of LLM inference.





\begin{figure*}[]
  \centering
  \begin{subfigure}[b]{0.565\textwidth}
      \includegraphics[width=\textwidth]{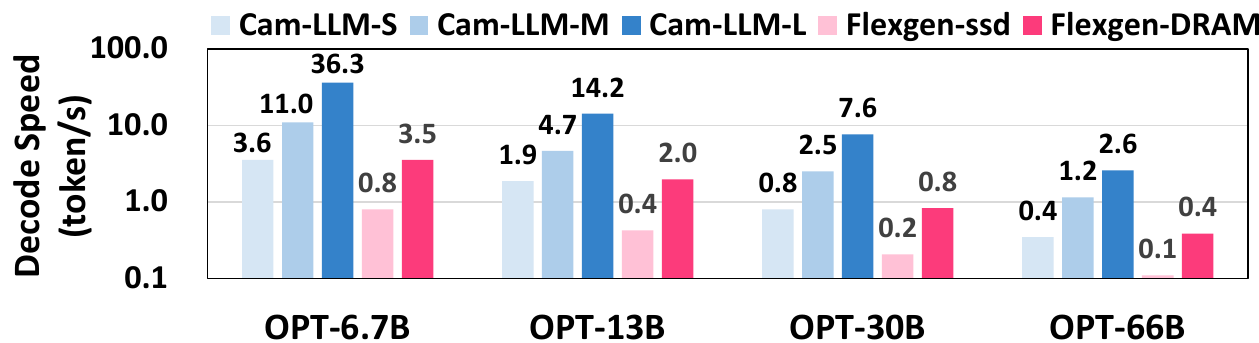}
      \caption{Performance of \xname~and Flexgen}
      \label{fig:exp_perf_opt}
  \end{subfigure}
  \begin{subfigure}[b]{0.38\textwidth}
      \includegraphics[width=\textwidth]{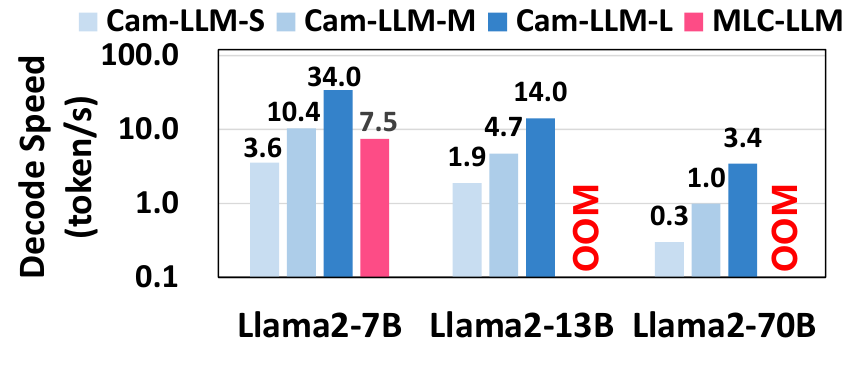}
      \vspace{-6mm}
      \caption{Performance of \xname~and MLC-LLM}
      \label{fig:exp_perf_llama}
  \end{subfigure}

  \vspace{-1mm}

  \caption{End-to-end performance of \xname~compared to (a) Flexgen and (b) MLC-LLM.}
  \label{fig:exp_performance}
\end{figure*}

\section{Experiment Setup}
\subsection{Hardware Configuration}
We compare the performance of \xname~with two mainstream LLM inference frameworks named Flexgen\cite{flexgen} and MLC-LLM\cite{mlc-llm}. 
Flexgen excels in offloading workloads to SSDs, while MLC-LLM is optimized for LLM inference on mobile devices. 
We present the hardware configurations of \xname~and the baselines as follows.

\begin{table}[!tbp]
  \centering
  \caption{Configurations of \xname}
  \label{table: 7_flash_settings}
  \vspace{-2mm}
  \resizebox{0.43\textwidth}{!}{
  \begin{tabular}{|c|ccc|}
    \hline
                      & \multicolumn{1}{c|}{\xname-S}                                    & \multicolumn{1}{c|}{\xname-M}                                    & \xname-L                                    \\ \hline
    Quantization      & \multicolumn{1}{c|}{8bit}                                          & \multicolumn{1}{c|}{8bit}                                          & 8bit                                          \\ \hline
    Channel           & \multicolumn{1}{c|}{8}                                             & \multicolumn{1}{c|}{16}                                            & 32                                            \\ \hline
    Chip              & \multicolumn{1}{c|}{2}                                             & \multicolumn{1}{c|}{4}                                             & 8                                             \\ \hline
    Flash Memory      & \multicolumn{3}{c|}{\begin{tabular}[c]{@{}c@{}}2 die   per chip, \\      2 plane and 1 compute core per die,\\      1000MT/s, 8 bit channel bus, \\      page size = 16KB\end{tabular}} \\ \hline
    Time   Parameters & \multicolumn{3}{c|}{tR = 30us}                                                                                                                                                          \\ \hline
    \end{tabular}
    }
  \end{table}

\textbf{\xname:~}
We use the well-known SSDsim simulator\cite{ssdsim} for Flash simulation and build a cycle-accurate simulator in C for NPU simulation.
For the Flash simulation, we expand the existing Read/Write command support in SSDsim to include Read Compute commands for in-storage computing. 
As shown in \autoref*{table: 7_flash_settings}, we implement three versions of \xname~and tailor each setting to different flash resources, named \xname-S, \xname-M, and \xname-L.
For the NPU, we utilize a 16x16 systolic array, capable of delivering 2 TOPS of computational power at a 1 GHz clock speed. 
The NPU is interfaced with LPDDR5X DRAM, which provides approximately 40GB/s of bandwidth. 
The DRAM is exclusively used for storing the KV cache, and a capacity of 700MB suffices for the needs of a 70B LLM under single batch inference. 
All test results are based on INT8 quantization, offline quantization techniques ensure model accuracy under this quantization level.
It should be noted that all quantization techniques are orthogonal to our architectural optimizations, \xname~will proportionally benefit from the development of more aggressive quantization strategies, such as 4/2 bit.

\begin{table}[tbp]
  \centering
  \caption{Configurations of baselines}
  \vspace{-2mm}
  \label{table: 7_baseline_settings}
  \resizebox{0.43\textwidth}{!}{
  \begin{tabular}{|c|cc|c|}
    \hline
                                                                          & \multicolumn{1}{c|}{Flexgen-SSD}                                          & Flexgen-DRAM                                         & MLC-LLM                                                                           \\ \hline
    Quantization                                                          & \multicolumn{1}{c|}{8bit}                                                 & 8bit                                                 & 4bit                                                                              \\ \hline
    Weight Offloading                                                     & \multicolumn{1}{c|}{SSD}                                                  & DRAM                                                 & DRAM                                                                              \\ \hline
    \begin{tabular}[c]{@{}c@{}}Hardware\\      Configuration\end{tabular} & \multicolumn{2}{c|}{\begin{tabular}[c]{@{}c@{}}AMD EPYC 7742 CPU,\\ A100 80G GPGPU,\\ Intel NVMe SSD,\\ 128GB DRAM\end{tabular}} & \begin{tabular}[c]{@{}c@{}}Qualcomm\\      Snapdragon\\      8 Gen 2\end{tabular} \\ \hline
    \end{tabular}
    }
  \vspace{-2mm}
\end{table}

\textbf{Flexgen:}
As shown in \autoref*{table: 7_baseline_settings}, We deploy Flexgen on a server and test its performance under different configurations. 
When using the Flexgen framework, we place both attention and activation on the GPU's HBM and place the model weights on the system DRAM (Flexgen-DRAM) and NVMe SSD (Flexgen-SSD) respectively. Since Flexgen only supports OPT, we only test its performance on the OPT series model.

\textbf{MLC-LLM:}
We test the performance of MLC-LLM on smartphones equipped with the Qualcomm Snapdragon 8 Gen 2 Soc. 
However, it is important to point out that the Llama2-7B model supported on MLC-LLM is based on 4-bit round-to-nearest (RTN) quantization. 



\subsection{Benchmarks}
We evaluate the performance of \xname~under on widely used LLMs such as OPT\cite{opt} and Llama2 \cite{llama2}, which have parameter sizes ranging from 6.7B to 70B. 
We conduct model accuracy experiments on popular datasets including HellaSwag\cite{hellaswag}, Arc\cite{arc}, and WinoGrande\cite{winogrande}. 
For the accuracy tests, we utilize the most advanced offline quantization framework, Smoothquant \cite{quant3_smoothquant}, to quantize model weights into INT8. Additionally, we construct flash error models of varying intensities using PyTorch and inject them into quantized model weight.



\subsection{\textcolor{black}{Area and Power Estimation}}
\textcolor{black}{
  We developed the computational core of \xname~using Verilog HDL and synthesized it through the Design Compiler utilizing the TSMC 65nm process technology node. As indicated in \autoref*{table: exp_area_power}, the primary contributors to overhead are input buffer and output buffer. These buffers serve the purpose of storing input and output vectors, with a combined capacity of 2KB, which suffices for the majority of applications. The area overhead for the compute core is 1.2\% and the power overhead is 4.5\%, both of which are within acceptable limits.}

\begin{table}[]
  \centering
  \caption{\textcolor{black}{Area and power overhead of compute core}}
  \vspace{-2mm}
  \label{table: exp_area_power}
    \resizebox{0.43\textwidth}{!}{
  \begin{tabular}{|l|l|l|}
    \hline
    \multicolumn{1}{|c|}{}         & Area (um$^2$) & Power (uw) \\ \hline
    Error   Correction Unit        & 496.4     & 0.4       \\ \hline
    PEs                            & 562.0     & 343.6     \\ \hline
    Input Buffer and Output Buffer & 58755.1   & 1591.7   \\ \hline
    Total Compute   Core           & 39813.5   & 1935.6   \\ \hline
    Overhead                       & 1.2\%        & 4.5\%        \\ \hline
  \end{tabular}
  }
  \vspace{-2mm}
\end{table}

\section{Experiment Result}


\begin{figure*}[!t]
  \centering
  \begin{subfigure}{0.32\linewidth}
      \centering
      \setlength{\abovecaptionskip}{0pt}
      \includegraphics[width=1.0\textwidth]{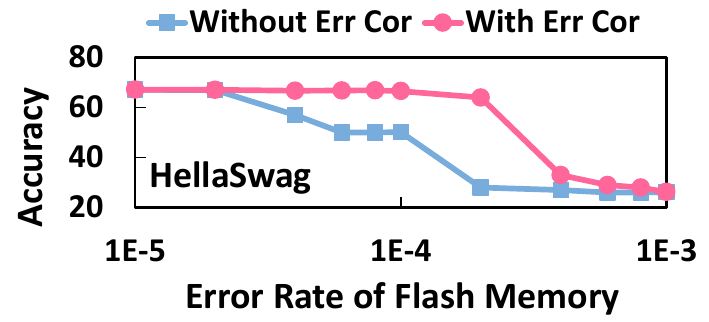}
      \caption{}
      \label{fig:flip_protect_1}
  \end{subfigure}
  \begin{subfigure}{0.32\linewidth}
      \centering
      \setlength{\abovecaptionskip}{0pt}
      \includegraphics[width=1.0\textwidth]{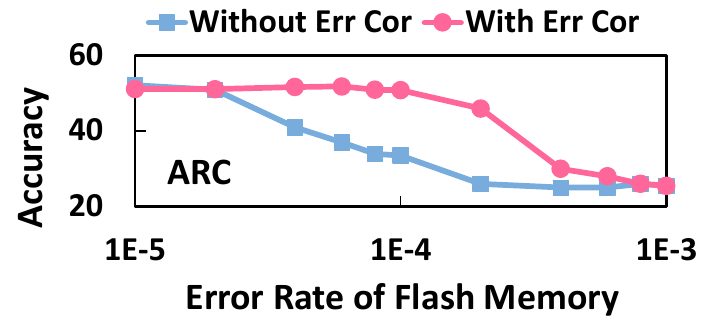}
      \caption{}
      \label{fig:flip_protect_2}
  \end{subfigure}
  \begin{subfigure}{0.32\linewidth}
    \centering
      \setlength{\abovecaptionskip}{0pt}
    \includegraphics[width=1.0\textwidth]{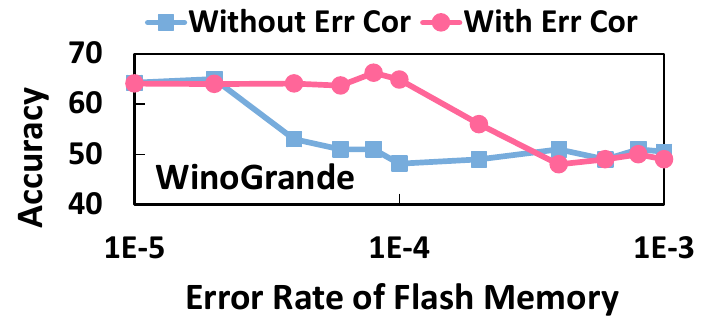}
    \caption{}
    \label{fig:flip_protect_3}
\end{subfigure}
    \vspace{-5pt}
  \caption{Accuracy evaluation of the proposed error correction mechanism.}
  \label{fig:exp_flip_protect}
    \vspace{-12pt}
\end{figure*}

\subsection{End-to-end performance}
This section presents the end-to-end decode speed of \xname~compared to the baselines.
\autoref*{fig:exp_perf_opt} shows the decode speed of \xname~and Flexgen for OPT models with varying parameter sizes.
\textbf{\xname-L} achieves rapid speed across all sizes of OPT models. 
On the OPT-66B model, its decode speed reaches 2.59 tokens/s, which is 22.1$\times$ faster than Flexgen-SSD and 6.8$\times$ faster than Flexgen-DRAM. 
On the OPT-6.7B model, the inference speed of \xname-L even reaches 36.34 tokens/s, which is 44.8$\times$ faster than Flexgen-SSD and 10.3$\times$ faster than Flexgen-DRAM.
\textbf{\xname-M}  achieved a speed comparable to Flexgen-DRAM across various tasks, with 10.96/4.68/2.50/1.15 tokens/s for the OPT-6.7B/13B/30B/66B models, respectively.
\textbf{\xname-S}, the configuration with minimal flash resources, attained a speed of 3.56 tokens per second on the OPT-6.7B model, achieving a smooth operational standard and outperforming Flexgen-SSD by 8.9$\times$.

\autoref*{fig:exp_perf_llama} illustrates the decoding speed of \xname~and MLC-LLM using Llama2 models with varying parameter sizes.
MLC-LLM places all LLM weights entirely in DRAM for inference, thus it is restricted to supporting only up to 7B parameters. 
On Llama2-13B and 70B, it encounters out-of-memory issues. 
When utilizing the Qualcomm 8 Gen 2 processor, MLC-LLM achieves a decoding speed of 7.58 tokens/s on Llama2-7B, which is higher than \xname-S's speed of 3.55 tokens/s. 
This performance disparity is due to MLC-LLM's use of 4-bit quantization, which results in model size reduction but significant precision loss compared to the 8-bit quantization used by \xname-S. 
Theoretically, employing 4-bit quantization in \xname-S as well  could improve the inference speed to match the MLC-LLM.

\begin{figure}[]
  \centering
  \begin{subfigure}{\linewidth}
      \centering
      \setlength{\abovecaptionskip}{-3mm}
      \includegraphics[width=1.0\textwidth]{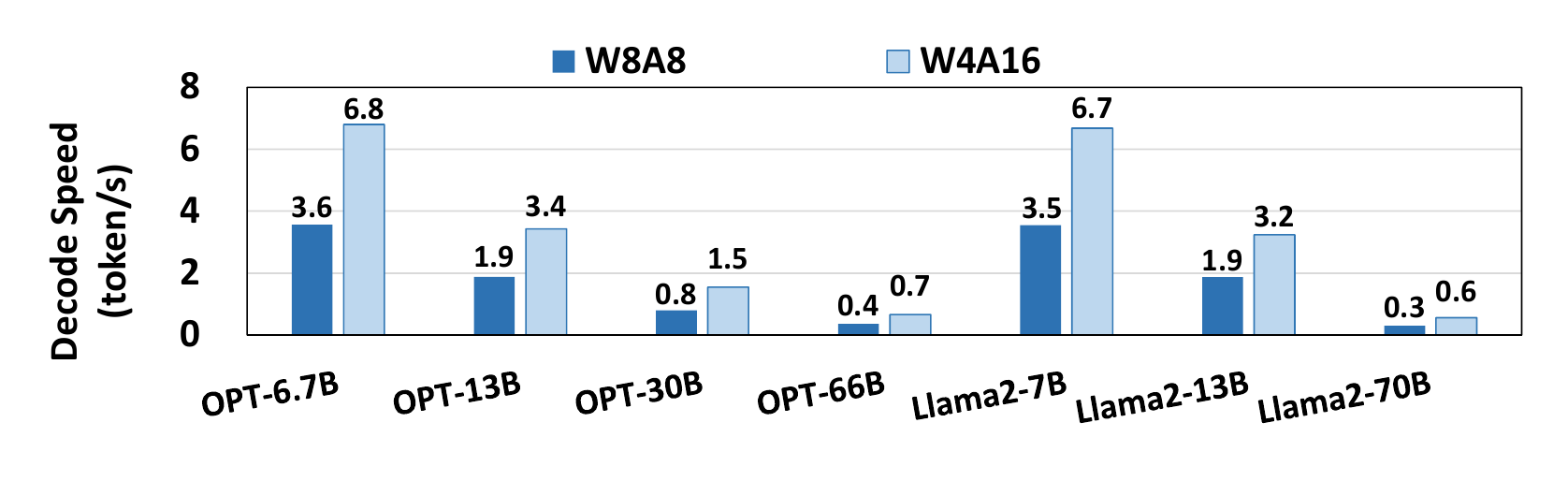}
      \caption{\textcolor{black}{Performance of \xname-S}}
      \label{fig:exp_4bit_S}
  \end{subfigure}
  \begin{subfigure}{\linewidth}
      \centering
    \setlength{\abovecaptionskip}{-3mm}
      \includegraphics[width=1.0\textwidth]{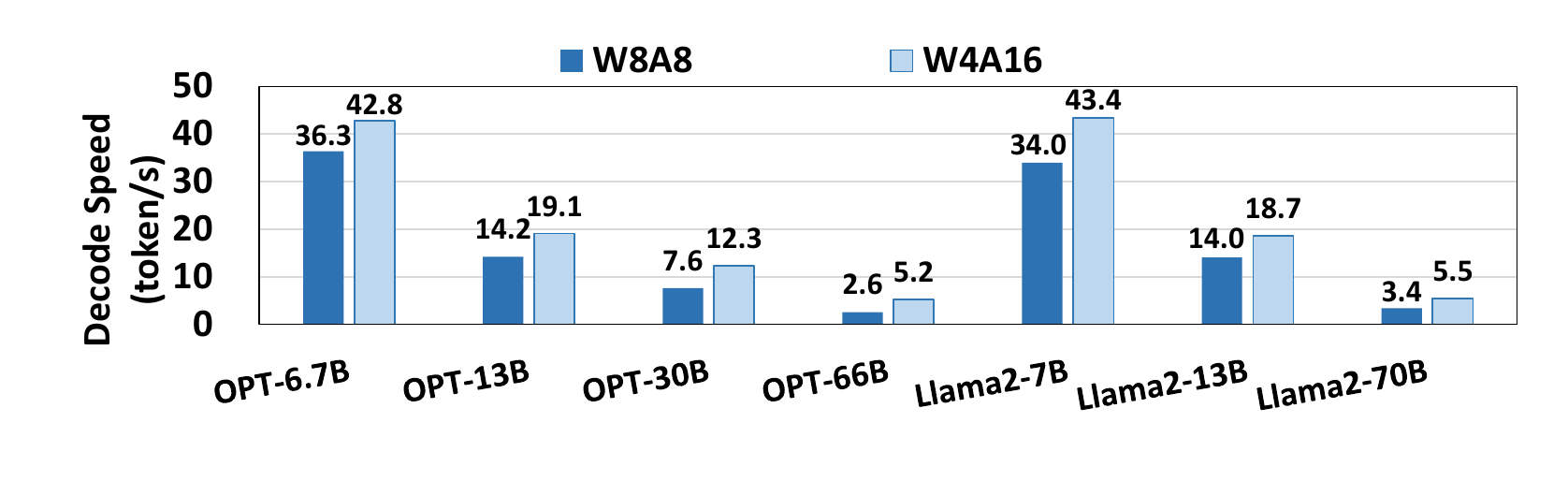}
      \caption{\textcolor{black}{Performance of \xname-L}}
      \label{fig:exp_power} 
  \end{subfigure}

  \vspace{-2mm}
  \caption{\textcolor{black}{Performance of \xname~under W4A16 quantization}}
  \label{fig:exp_4bit}
  \vspace{-2mm}
\end{figure}

\subsection{\textcolor{black}{Performance under 4 bit Quantization}}
\textcolor{black}{To demostrate how our \xname~architecture can benifit from lower bit quantization techniques, we conducted experiment evaluating the performance of \xname~with 4 bit quantization for weight and 16 bit quantization for activation (W4A16). In comparison to the defaut 8 bit quantization (W8A8), this W4A16 quantization of model weight requires less bandwidth. \autoref*{fig:exp_4bit} presents a comparison of W8A8 and W4A16 quantizations across both \xname-S and \xname-L configurations. On average, W4A16 quantization yields performance improvements of 85.3\% for \xname-S and 47.9\% for \xname-L. Additionally, we observed that larger performance improvements occur in larger LLMs, which face more severe memory constraints when loading model weights. Reducing the bit-width of model weights significantly boosts their performance.}

\subsection{Ablation Study}
This section analyzes the performance improvement with the two main techniques of \xname, \ie, read request slicing and hardware-aware tiling. \textcolor{black}{This section also test the affect of different tile sizes}

\begin{figure}[t]
  \centering
  \begin{subfigure}{\linewidth}
      \centering
      \setlength{\abovecaptionskip}{-2mm}
      \includegraphics[width=1.0\textwidth]{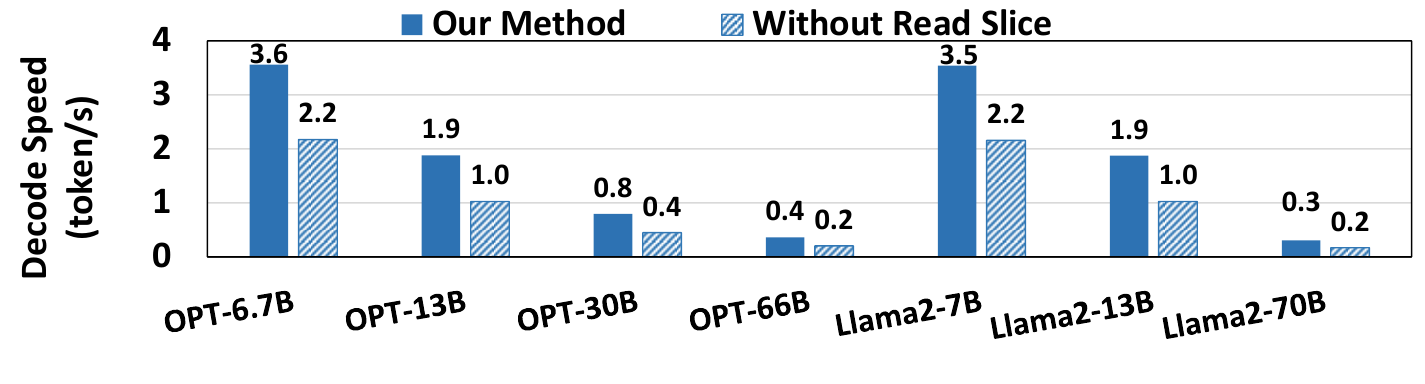}
      \caption{Decode Speed comparison}
      \label{fig:exp_no_read_slice_speed}
  \end{subfigure}
  
  \vspace{2mm}

  \begin{subfigure}{\linewidth}
      \centering
      \setlength{\abovecaptionskip}{-1mm}
      \includegraphics[width=1.0\textwidth]{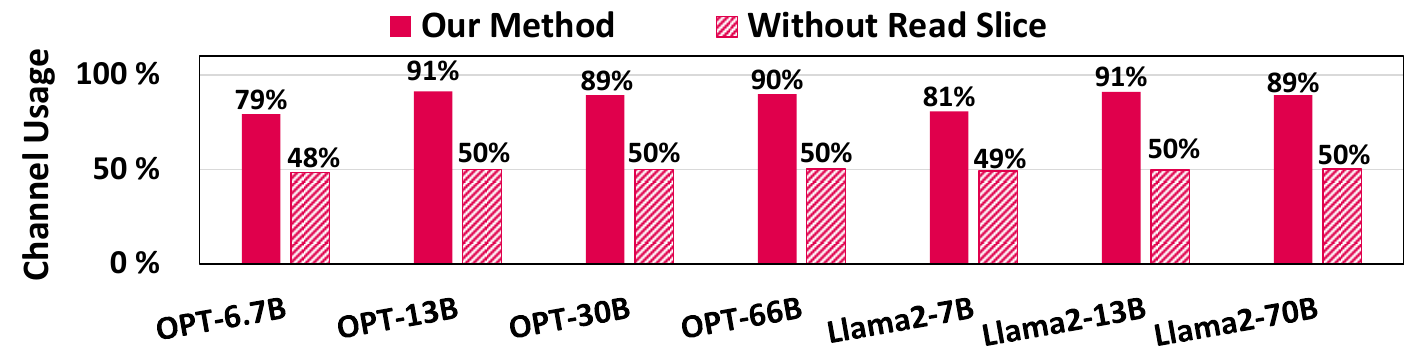}
      \caption{Channel Usage comparison}
      \label{fig:exp_no_read_slice_usage}
  \end{subfigure}
  
  \vspace{-2mm}
  \caption{Ablation study of the read request slice feature.}
  \label{fig:exp_no_read_slice}
\end{figure}

For read request slicing, we use \xname-S configuration and compare the full-featured \xname~with a simplified version without this feature. 
In the simplified version, read requests are sent in a continuous, unsegmented manner. 
\autoref*{fig:exp_no_read_slice} shows the results for decoding speed and channel utilization on different LLM configurations. 
The results show that the read request slicing achieves a speed up ranging from 1.6x to 1.8x and increases bandwidth utilization by 31.6\% to 41.4\%. 
This improvement is due to the sliced read requests not blocking the execution of read compute requests.
Read request data will only occupy the channel slice by slice during the bubbles in read compute requests.

\begin{figure}[!t]
  \centering
  \begin{subfigure}{\linewidth}
      \centering
      \setlength{\abovecaptionskip}{0mm}
      \includegraphics[width=1.0\textwidth]{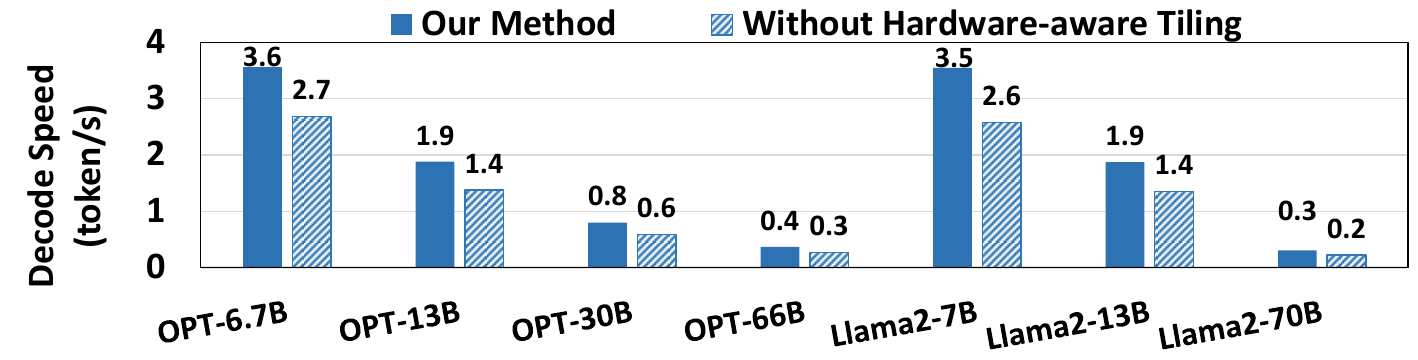}
      \caption{Decode Speed Comparison}
      \label{fig:exp_no_distribution_speed}
  \end{subfigure}

  \vspace{2mm}
  
  \begin{subfigure}{\linewidth}
      \centering
      \setlength{\abovecaptionskip}{0mm}
      \includegraphics[width=1.0\textwidth]{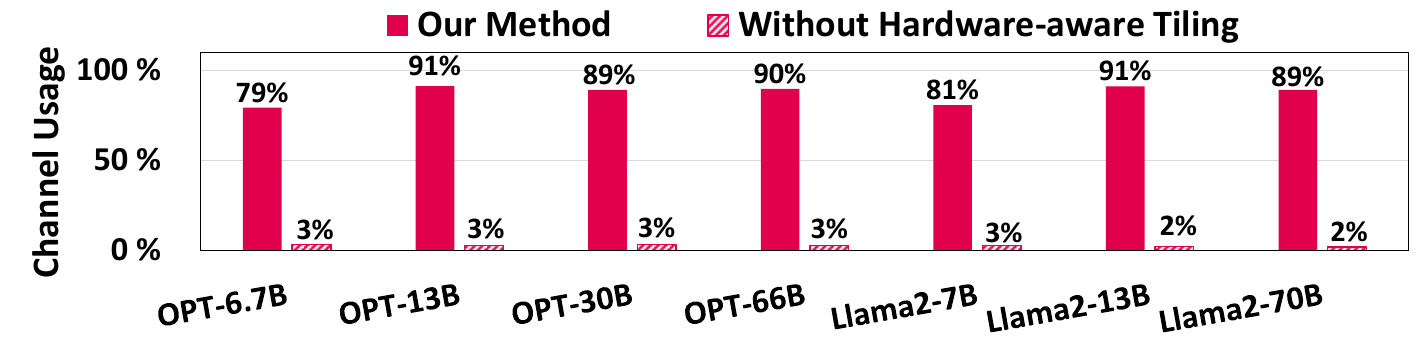}
      \caption{Channel Usage Comaprison}
      \label{fig:exp_no_distribution_usage} 
  \end{subfigure}

  \vspace{-2mm}
  \caption{Ablation study of the hardware-aware tiling feature.}
  \label{fig:exp_no_distribution}
\end{figure}

For the hardware-aware tiling, we again use the \xname-S configuration and compare the full-featured \xname-S with a version that did not employ the tiling. 
Without tiling, all operations involving the multiplication of weight matrices and input vectors are completed by Flash, with no offloading to NPU. 
\autoref*{fig:exp_no_distribution} displays the differences in inference speed and channel utilization on different LLMs. 
The results show that hardware-aware tiling can accelerate LLM inference by 1.3x to 1.4x and increase channel utilization by 76.2\% to 88.9\%. 
This is because the tiling strategy utilizes idle channel bandwidth to transfer some weights to the NPU for computation, thereby reducing the workload of on-die processing on the flash.

\begin{figure}[!t]
  \centering
  \setlength{\abovecaptionskip}{-1mm}
  \includegraphics[width=0.5\textwidth]{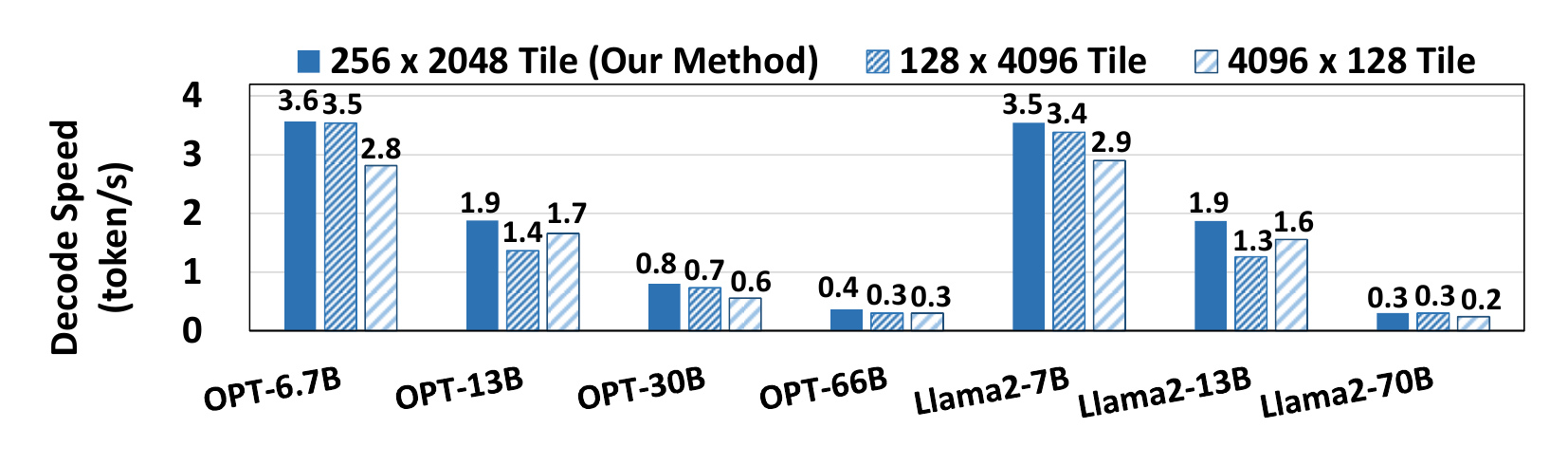}
  \caption{\textcolor{black}{Performance of \xname~under varied tile sizes.}}
  \label{fig:exp_tile}
  \vspace{-4mm}
\end{figure}

\textcolor{black}{We conducted further experiments to assess how tile sizes influence the performance of \xname. In these experiments, We facilitated collaborative operation between the NPU and flash to handle GEMV operations, and explored the effects of multiple sub-optimal tile sizes on performance. We selected the \xname-S configuration with an theoretically optimal tile size of $256 \times 2048$ (Height $\times$ Width). Additionally, we evaluated the performance of two other configurations with tile sizes of $128 \times 4096$ and $4096 \times 128$. As illustrated in \autoref*{fig:exp_tile}, the optimal tile size of $256 \times 2048$ outperforms the $128 \times 4096$ configuration by 17.5\% and the $4096 \times 128$ configuration by 24.7\%.}

\subsection{On-Die Error Correction Mechanism}
We evaluate the sensitivity of the OPT-6.7B model to flash memory errors and the effectiveness of the proposed on-die error correction mechanism on three datasets: HellaSwag, ARC, and WinoGrande. 
As shown in \autoref*{fig:exp_flip_protect}, without the Error Correction Unit, the accuracy of the LLM begins to significantly decrease when the error rate reaches $1\times10^{-5}$. When the error rate further achieves $2\times10^{-4}$, the accuracy falls to about 40\% of its original level, making the output unreliable. 
On the other hand, with \xname's on-die Error Correction Unit enabled, the LLM can maintain 92\%$\sim$95\% of its original accuracy at an error rate of $2\times10^{-4}$, providing 2.3$\times$ protection capabilities compared to baseline. 

However, the protection capability of \xname~also has its limits.
When the error rate exceeds $8\times10^{-4}$, the accuracy of LLM still drops to an unusable level even with the Error Correction Unit enabled. 
This is because the proposed error correction mechanism only protects outliers from being changed and prevents those trivial values below the threshold from flipping into fake outliers.
While it offers no protection for intermediate and small values that do not reach the threshold before and after potential bit flips.
It is the extensive flipping of these intermediate and small values that leads to the significant degradation in LLM accuracy.

\subsection{Scalability Study}
To assess the scalability of \xname, we evaluate the impact of flash channel numbers and chip numbers on the decode speed of \xname~on OPT-6.7B/13B/30B.

\begin{figure}[]
  \centering
  \begin{subfigure}{0.49\linewidth}
      \centering
      \setlength{\abovecaptionskip}{-1pt}
      \includegraphics[width=1.0\textwidth]{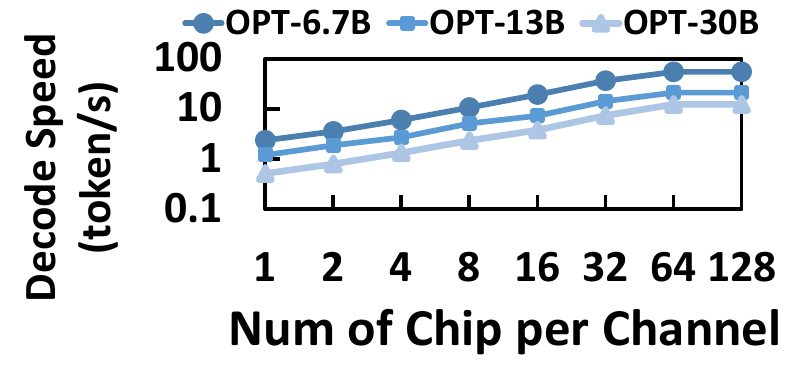}
      \caption{}
      \label{fig:exp_scale_1}
  \end{subfigure}
  \begin{subfigure}{0.49\linewidth}
      \centering
      \setlength{\abovecaptionskip}{-1pt}
      \includegraphics[width=1.0\textwidth]{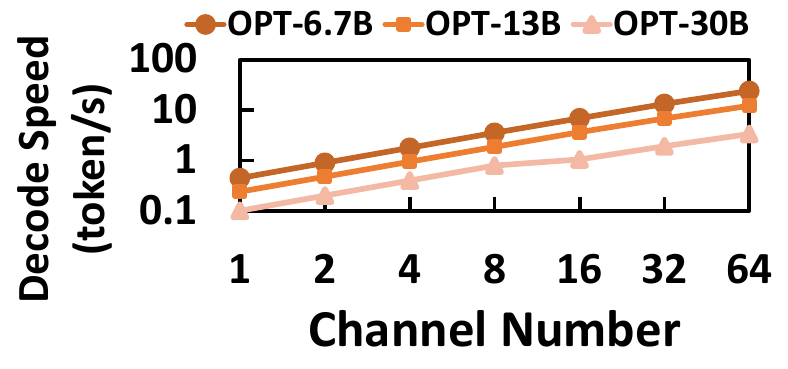}
      \caption{}
      \label{fig:exp_scale_2}
  \end{subfigure}
  \begin{subfigure}{0.49\linewidth}
      \centering
      \setlength{\abovecaptionskip}{-1pt}
      \includegraphics[width=1.0\textwidth]{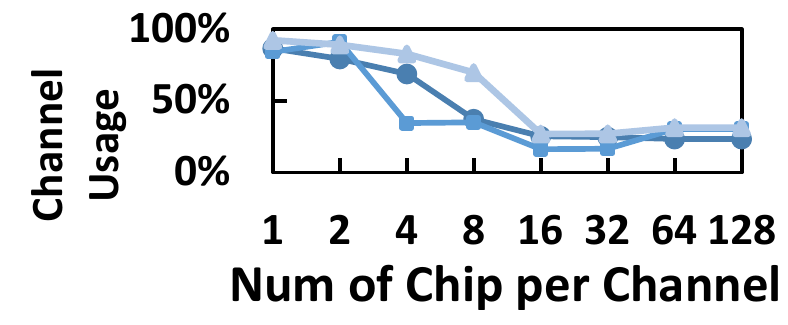}
      \caption{}
      \label{fig:exp_scale_3}
  \end{subfigure} 
  \begin{subfigure}{0.49\linewidth}
      \centering
      \setlength{\abovecaptionskip}{-1pt}
      \includegraphics[width=1.0\textwidth]{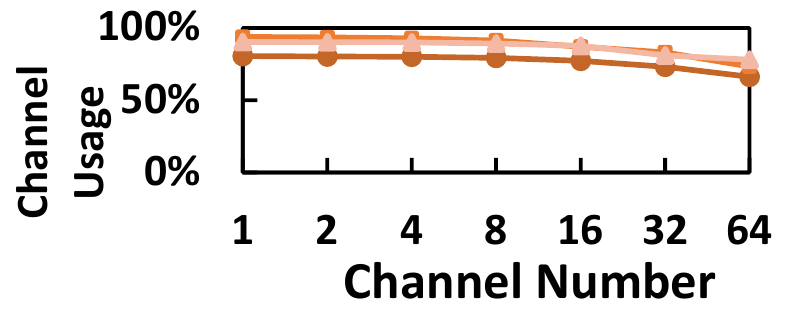}
      \caption{}
      \label{fig:exp_scale_4}
  \end{subfigure}
  
  \vspace{-2mm}
  \caption{Scalebility evaluation of channel and chip number.}
  \vspace{-4mm}

  \label{fig:exp_scale}
\end{figure}

When examining the effect of the number of chips per channel, we set the number of channels to 8 and increased the number of chips mounted on each channel from 1 to 128. 
As shown in \ref*{fig:exp_scale}, the decode speed increases rapidly with the number of chips at the beginning, but the growth gradually slows down. 
This slowdown is attributed to the excessive number of chips, which prevents the model weight from being effectively distributed across all chips. 
Even with an increased number of chips, many remained idle, yielding no performance gains. 
As depicted in \ref*{fig:exp_scale}, the utilization of the channels noticeably decreases when there are too many chips on a channel. 
This decrease is due to an increase in the number of compute cores, leading to more on-die computational power in the flash.
Consequently, more computations are allocated to the on-die resources, reducing the computations assigned to the NPU, and thus channels are no longer required to transmit the weight matrix, lowering their utilization.

During the evaluation of how channel count influences performance, we set the number of chips per channel to 4 and incrementally increase the number of channels from 1 to 64. 
As shown in \autoref*{fig:exp_scale}, \xname's performance steadily increases with the number of channels while the channel utilization slowly declines within the test range. 
This demonstrates good scalability of \xname~performance with increasing channel numbers. 
As the maximum number of flash channels increases, the performance of \xname~is expected to improve further. 
However, it must be noted that since \xname~employs chiplet technology to link the NPU and Flash, the number of channels is also constrained by the chiplet fabrication technology and the geometric dimensions.
Moreover, in order to accommodate a greater number of channels, the buffer on the NPU needs to be proportionally expanded to hold a larger amount of data from the flash. 
However, this enlargement of the buffer will result in increased area consumption.


\subsection{Data Transfer and Power Consumption}
\begin{figure}[t]
  \centering
  \begin{subfigure}{\linewidth}
      \centering
      \setlength{\abovecaptionskip}{-1pt}
      \includegraphics[width=1.0\textwidth]{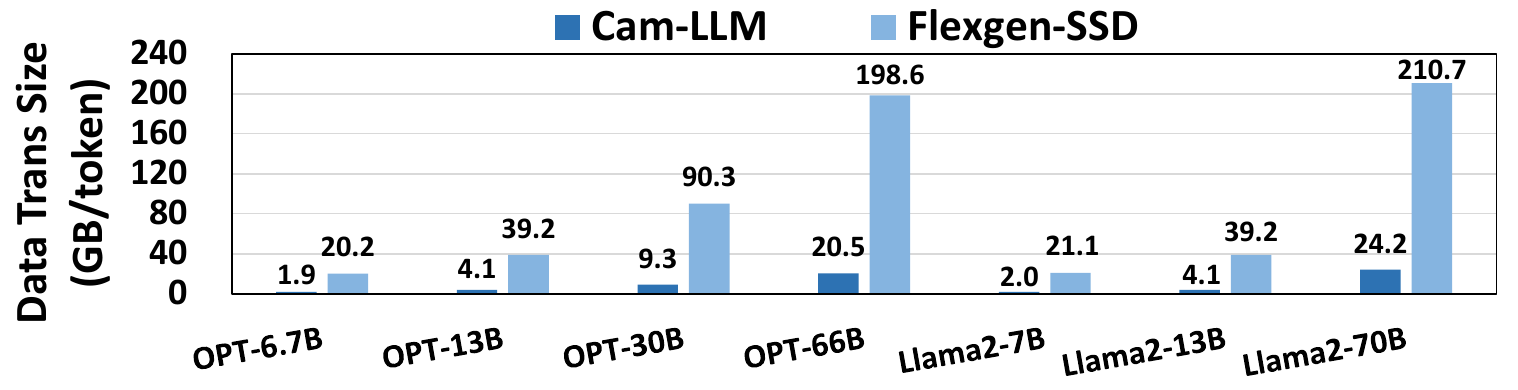}
      \caption{Data transfer size comparison}
      \label{fig:exp_data_trans}
  \end{subfigure}
  \begin{subfigure}{\linewidth}
      \centering
    \setlength{\abovecaptionskip}{-1pt}
      \includegraphics[width=1.0\textwidth]{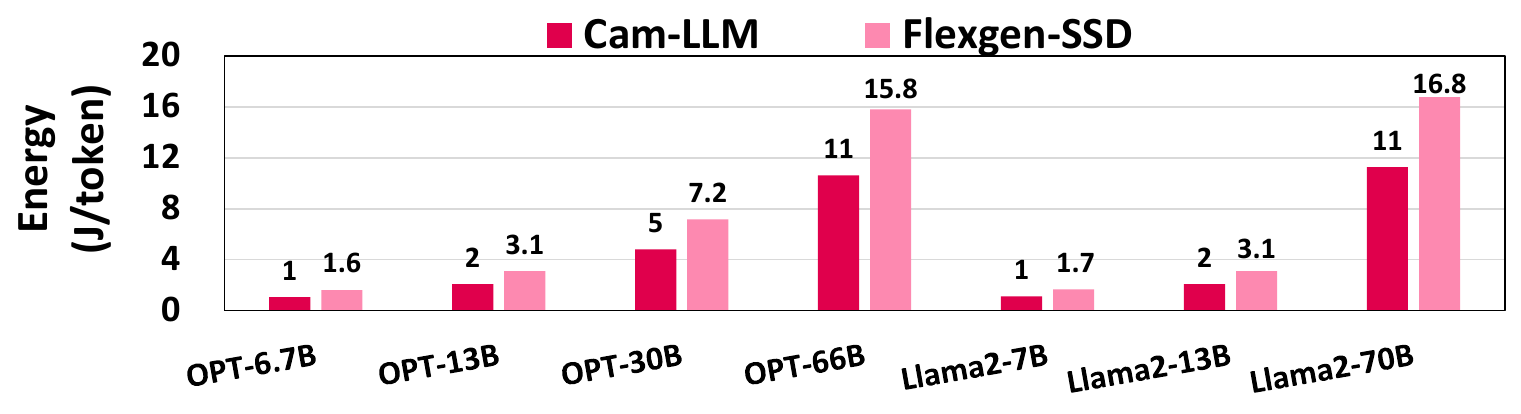}
      \caption{Power consumption comparison}
      \label{fig:exp_power} 
  \end{subfigure}
  
  \vspace{-2mm}
  \caption{Data transfer size comparison and power consumption comparison of \xname~-S and Flexgen-SSD.}
  \label{fig:exp_data_and_power}
    \vspace{-12pt}
\end{figure}

We evaluate the data transfer size and energy consumption of one token during inference for \xname-S and Flexgen-SSD across various models. 
The results are shown in \autoref*{fig:exp_data_and_power}. 
\xname-S transfers 9.7$\times$ to 11.6$\times$ less data Flexgen-SSD.
This substantial reduction can be attributed to two primary factors. 
Firstly, the NPU within \xname~has direct access to the data stored in Flash, which eliminates the need for transferring data from Flash to DRAM. 
Secondly, the Flash in \xname~possesses in-storage computing capabilities, allowing for data to be processed and its size reduced within the flash die before transmission.
Furthermore, the total energy consumed by \xname-S for data transfer per token is only 67\% of that used by Flexgen-SSD. 
This reduction in energy consumption is primarily due to the decreased size of data transfer and the incorporation of chiplet technology, which reduces the overhead associated with data transmission between NPU and Flash.


\subsection{\textcolor{black}{Cost Analysis}}
\textcolor{black}{
This section compares the cost of \xname~ with traditional architecture. To support the inference of 70B LLM under 8 bit quantization, a minimum storage of 80 GB is needed. Traditional architecture uses DRAM to store all model weights and KV cache while \xname~ leverages considerably cheaper flash memory to store model weights, reserving DRAM solely for KV cache. As shown in \autoref*{table: exp_cost}, \xname~offers a cost advantage, being \$150.01 cheaper than the traditional architecture. 
}

\textcolor{black}{
However, it is essential to acknowledge two additional costs associated with \xname~that are not captured in \autoref*{table: exp_cost}. Firstly, the integration of new logic into flash die incurs extra costs. Nonetheless, this additional logic is relatively simple and area overhead is minimal, redering the associated cost negligible. Secondly, the adoption of chiplet technology in \xname~also contributes to overall costs. While specific open-source data on chiplet technology costs are unavailable, recent research on the cost model of chiplets~\cite{chiplet_cost} suggests that expenses related to the die-to-die interface and packaging should be less than 15\% of the raw chip costs, estimated not to exceed \$100 in the case of \xname.}

\section{Related Work}
\textbf{LLM acceleration.} There exists a number of prior works to accelerate LLMs. 
Some works\cite{mlsys1_slora,mlsys2_pagedattention, mlsys3_alg_to_system,mlsys4_Powerinfer,mlsys5_punica,mlsys6_orca} focus on huge batch size  LLM inference with substantial hardware resources, while others aim to maximize the use of limited available GPU resources\cite{limit1_flash1,limit2_flash2}. 
Additional studies take advantage of the sparsity of attention matrices to make software optimization \cite{sparse1_spatten,sparse4_longformer,sparse5_bigbird}, design new architectures \cite{sparse3_sanger,sparse2_a3,sparse6_hardware_acc,sparse7_co_design,sparse8_salo,sparse9_dysta} and implement FPGA deployment \cite{fpga1_pruning,fpga2_ftrans,fpga3_co,fpga4_cgra,fpga5_compression}.
However, none of these approaches have the potential to enable the inference of a 70B LLM on smartphones due to DRAM size limitations. 
To address DRAM constraints, some works have proposed offloading strategies to flash-based SSDs \cite{flexgen,DeepSpeed}, though limited bandwidth slows down the inference speed. The "LLM in a Flash"\cite{llm_in_a_flash} method attempts to overcome this by exploiting the sparsity of LLMs to reduce data transfer volumes; however, it still requires substantial DRAM and fails to utilize the inherent parallelism of flash memory effectively.
Furthermore, numerous quantization-based approaches \cite{quant1_olive,quant2_awq,quant3_smoothquant,quant4_rptq,quant5_fp4,quant6_gptq,quant7_qlora} can also help with the deployment of LLM on edge devices since they can reduce the total weight size. 
Notably, these quantization methods are all orthogonal to \xname.

\textbf{In-Storage processing.} In-storage computing (ISC) has been widely used for many applications such as data search and analytics \cite{data1_smart,data2_summarizer,data3_deepstore,data4_catalina,data5_assasin}, DNN \cite{dnn1_thrifty,dnn2_behemoth,recssd2021,ecssd}, biological information processing \cite{bio1_spectrum,genstore2022}  and graph processing \cite{graph1_grafboost,graph2_graphssd}. However, all of these works add the computing logic near the controller, failing to explore the potential of processing on flash dies.

\textbf{On-die Processing.} Only a few works have proposed the design of adding logic on flash dies. Deepstore \cite{data3_deepstore} equipped flash die with a systolic array for query searching, while the logic is too complex to be feasible. Optimstore \cite{optimstore} and \cite{beacongnn} adopt on-die logic to accelerate DNN training and large-scale GNN, respectively. However, both of these designs ignore the importance of on-die error correction, making them unusable for LLM inference tasks. Besides, the high reduction ratio of LLM single-batch inference leads to server low flash channel utilization, making them not suitable for this task.

\begin{table}[!t]
    \centering
    \caption{\textcolor{black}{Cost of \xname~and traditional architecture to support inference of 70B LLM under 8 bit quantization.}}
    \label{table: exp_cost}
    \vspace{-1mm}
    \resizebox{0.43\textwidth}{!}{
    \begin{tabular}{|c|cc|cc|}
    \hline
    \multirow{2}{*}{} & \multicolumn{2}{c|}{~\xname}          & \multicolumn{2}{c|}{Traditional Architecture} \\ \cline{2-5} 
                      & \multicolumn{1}{c|}{Count} & Cost (\$) & \multicolumn{1}{c|}{Count}     & Cost (\$)    \\ \hline
    DRAM (GB)         & \multicolumn{1}{c|}{2}     & 4.87      & \multicolumn{1}{c|}{80}        & 194.68       \\ \hline
    Flash (GB)        & \multicolumn{1}{c|}{80}    & 38.80     & \multicolumn{1}{c|}{0}         & 0.00         \\ \hline
    Total Price       & \multicolumn{1}{c|}{N/A}   & 43.67     & \multicolumn{1}{c|}{N/A}       & 194.68       \\ \hline
    \end{tabular}
    }
    \vspace{-2mm}
  \end{table}

\section{Conclusion}
We propose \xname, a chiplet-based architecture consists of an NPU and a flash to enable the on-device inference of 70B LLMs.
The chiplet design and the direct access between the NPU and the flash ensure minimal energy consumption during data movement.
The flash is provided with on-die processing capabilities and works collaboratively with the NPU to handle GeMV multiplication through the proposed hardware-aware tiling strategy. To deal with the high error rate of flash memory, we design a lightweight on-die Error Correction Unit in the flash to guarantee reliability. \xname~can conduct inference of 70B LLM at a speed of 3.4 token/s, which is over 22$\times$ faster than the state-of-the-art flash-offloading framework.

\section{Acknowledgment}

This work is partially supported by the National Key R\&D Program of China(under Grants 2023YFB4502200, 2023YFB4502702), the NSF of China(under Grants U20A20227, U22A2028, 61925208, 62302482, 62222214, 62341411, 62302478, 62102398, 62102399, 62302483, 62302480, 62202453, 62332021, 62472007), Strategic Priority Research Program of the Chinese Academy of Sciences, (under Grants No.XDB0660200, XDB0660201, XDB0660202), CAS Project for Young Scientists in Basic Research(YSBR-029), Youth Innovation Promotion Association CAS and Xplore Prize, State Key Lab of Processors, Institute of Computing Technology, CAS(under Grant CARCH6102).






\end{document}